Roadmap

# The 2024 Active Metamaterials Roadmap

Simon A. Pope[1,18,19], Diane J. Roth[2,18], Aakash Bansal[3,18], Mostafa Mousa[4], Ashkan Rezanezad[4], Antonio E. Forte[4], Geoff. R. Nash[5], Lawrence Singleton[6], Felix Langfeldt[6], Jordan Cheer[6], Stephen Henthorn[7], Ian R. Hooper[8], Euan Hendry[8], Alex W. Powell[8], Anton Souslov[9], Eric Plum[10], Kai Sun[11], C. H. de Groot[12], Otto L. Muskens[11], Joe Shields[8], Carlota Ruiz De Galarreta[8,13], C. David Wright[8], Coskun Kocabas[14], M. Said Ergoktas[14], Jianling Xiao[15], Sebastian A. Schulz[16], Andrea Di Falco[15], Alexey V. Krasavin[2], Anatoly V. Zayats[2], Emanuele Galiffi[17]

[1] Department of Automatic Control and Systems Engineering, The University of Sheffield, Amy Johnson Building, Portobello Street, Sheffield, S1 3JD, UK

[2] Department of Physics and London Centre for Nanotechnology, King's College London, Strand, WC2R 2LS, UK

[3] Wolfson School of Mechanical, Electrical, and Manufacturing Engineering, Loughborough University, UK

[4] Department of Engineering, King's College London, London, WC2R 2LS, UK

[5] Natural Sciences, Harrison Building, North Park Road, University of Exeter, Exeter, UK

[6] Institute of Sound & Vibration Research, University of Southampton, University Road, Southampton, UK

[7] Department of Electronic and Electrical Engineering, University of Sheffield, Sheffield, UK

[8] Centre for Metamaterial Research and Innovation, University of Exeter, Stoker Road, Exeter, EX4 4QL, UK

[9] T.C.M. Group, Cavendish Laboratory, University of Cambridge, Cambridge CB3 0HE, UK

[10] Optoelectronics Research Centre and Centre for Photonic Metamaterials, University of Southampton, Highfield, Southampton SO17 1BJ, UK.

[11] School of Physics and Astronomy, University of Southampton, Southampton UK

[12] Electronics and Computer Science, University of Southampton, Southampton UK

[13] Institute of Materials Science of Barcelona, Universitat Autònoma de Barcelona, 08193 Bellaterra, Barcelona, Spain

[14] Department of Materials, National Graphene Institute, University of Manchester, UK

[15] School of Physics and Astronomy, University of St Andrews, North Haugh, St. Andrews, Fife, KY16 9SS, UK

[16] Department of Electrical and Computer Engineering, University of Waterloo, 200 University Avenue West, Waterloo, ON N2L 3G1, Canada

[17] Photonics Initiative, Advanced Science Research Center, City University of New York (85 St. Nicholas Terrace, 10027, New York, NY, USA)

[18] Guest editors of the Roadmap.

[19] Author to whom any correspondence should be addressed.

E-mails: s.a.pope@sheffield.ac.uk, diane.roth@kcl.ac.uk, a.bansal@lboro.ac.uk

# Abstract

Active metamaterials are engineered structures that possess novel properties that can be changed after the point of manufacture. Their novel properties arise predominantly from their physical structure, as opposed to their chemical composition and can be changed through means such as direct energy addition into wave paths, or physically changing/morphing the structure in response to both a user or environmental input. Active metamaterials are currently of wide interest to the physics community and encompass a range of sub-domains in applied physics (e.g. photonic, microwave, acoustic, mechanical, etc.). They possess the potential to provide solutions that are more suitable to specific applications, or which allow novel properties to be produced which cannot be achieved with passive metamaterials, such as time-varying or gain enhancement effects. They have the potential to help solve some of the important current and future problems faced by the advancement of modern society, such as achieving net-zero, sustainability, healthcare and equality goals. Despite their huge potential, the added complexity of their design and operation, compared to passive metamaterials creates challenges to the advancement of the field, particularly beyond theoretical and lab-based experiments. This roadmap brings together experts in all types of active metamaterials and across a wide range of areas of applied physics. The objective is to provide an overview of the current state of the art and the associated current/future challenges, with the hope that the required advances identified create a roadmap for the future advancement and application of this field.

## Contents

# 1 -   Introduction

Simon A. Pope[1], Diane Roth[2] and Aakash Bansal[3]

[1] Department of Automatic Control and Systems Engineering, The University of Sheffield, Amy Johnson Building, Portobello Street, Sheffield, S1 3JD, UK

s.a.pope@sheffield.ac.uk

[2] Department of Physics and London Centre for Nanotechnology, King's College London, Strand, WC2R 2LS, UK

diane.roth@kcl.ac.uk

[3] Wolfson School of Mechanical, Electrical, and Manufacturing Engineering, Loughborough University, UK

a.bansal@lboro.ac.uk

Metamaterials are engineered structures that possess novel properties that arise predominantly from their physical structure, as opposed to their chemical composition. While the exact definition of "active" metamaterials is open for debate, the definition adopted here is that active metamaterials extend the concept of a metamaterial such that its engineered structure or its response can be changed, in some manner, after manufacture. This definition is broader than what might be classed as "pure active" metamaterials in which energy is added directly into the path of wave propagation in the structure. It also includes: "tuneable" metamaterials in which the novel properties change in response to a user's input; "adaptive" metamaterials in which the novel properties change/evolve in response to a stimulus from the environment; "reconfigurable" metamaterials in which the structure can be physically changed (this also includes "shape morphing" metamaterials); "programmable" metamaterials in which the novel properties can be programmed into a standard design which has a structure with a number of discrete states. In some instances a specific metamaterial might encompass more than one of these definitions and/or also cover several physical domains, such as electromagnetic (photonic, microwave, etc.) and mechanical (e.g. acoustic, structural, etc.) metamaterials. This inherently interdisciplinary roadmap covers a broad spectrum of active metamaterials, and focuses on synergies and common problems associated with the different physical domains and design types.

"Active" designs provide a powerful approach, through which metamaterials can provide new types of novel responses, in addition to making existing designs more applicable to certain applications. While the potential of active metamaterials is clear and enticing for academia and industry, there are significant challenges which need to be overcome for the practical realisation and adoption of the technology. Within this roadmap, we have tried to draw on the expertise and experience across this broadest spectrum, to highlight the historical perspective, current status, future perspectives and significant challenges. It is clear from the roadmap that the field of active metamaterials is rapidly developing, but the increased complexity in terms of design and implementation, such as the requirements for energy sources, some form of actuation and multi-material designs feeds through to associated problems with modelling, simulation and manufacture. A number of cross domain challenges are highlighted, including achieving practically suitable actuation for the scales and array sizes required and manufacturing arrays of meta-atoms without (or in some case with) irregularities and inconsistencies. There are also specific challenges in each domain or application area, such as small sizes and addressing individual meta-atoms without crosstalk. The use of machine learning and improvements in modelling and simulation are identified across several domains as routes to providing improved designs and actuation routines to achieve the desired performance for these complex systems. By addressing these the interdisciplinary field of active metamaterials can start to address some of the significant challenges faced by industry and society in general, such as achieving net-zero, sustainability, healthcare and equality goals.

## 2 -   Shape-Morphing Mechanical Metamaterial

Mostafa Mousa, Ashkan Rezanejad and Antonio E. Forte
Department of Engineering, King's College London, London, WC2R 2LS, UK
mostafa.mousa@kcl.ac.uk,  ashkan.rezanejad@kcl.ac.uk,  antonio.forte@kcl.ac.uk.

**Status**

For more than two decades, the ability to control shape has been a sought-after feature in mechanical metamaterials, as it enables applications across fields, spanning from shape mimicking [1, 2], soft robotics [3], reconfigurable structures [4] to flexible electronics [5].

In particular, morphing metamaterials offer features such as reconfigurability, adaptability, and flexibility with minimal need for external controls. Notably, these advantages might not need a continuous energy supply and can be achieved by harnessing kirigami/origami approaches, multi-stability principles, and more. Kirigami, is an ancient Japanese art which consists in embedding arrays of cuts onto sheets of material. Depending on the geometric parameters of the cuts [1] and the type of deformation applied to the material [6], these systems are able to morph into 2D or 3D shapes. Because of their easy fabrication, programmability and low cost, kirigami metamaterials have been used in the design of flexible and wearable electronics [7], actuators [1] and soft robots [8]. Similarly, Origami utilizes folds instead of cuts to create 3D shapes and structures from sheets of material. A unique characteristic of multi-stable structures is that they are reconfigurable post-fabrication. This differs from classical pre-programmed solutions, where the material is usually designed to achieve one shape, stiffness, or functionality, upon activation. Additionally, cells within a material can be independently triggered to switch from one state to another; a common example is snap-through/back arches and membranes, which are bistable and can transition between two stable states via different external stimuli.

These approaches aim to enhance mechanical systems and expand their design space. In summary, the key features that characterise shape-morphing mechanical metamaterials are (i) reconfigurability, defined as the material's ability to change shape into various forms based on different inputs, (ii) reversibility, in which the material can revert to its original shape after morphing, (iii) power independence, indicating that the material does not require a power supply to maintain its reconfigured form. Among these, we believe that reconfigurability is the cornerstone of shape morphing metamaterials, and a feature which is yet to be fully explored and utilised. Hence, in this section, we discuss the challenges and opportunities to further develop reconfigurable and shape-morphing mechanical metamaterials.

**Current and Future Challenges**

Recent strategies have attempted to achieve reconfigurability by developing deformable mechanical metamaterials which are responsive to specific external forces (e.g., pneumatic, tensile). The primary challenge lies in achieving an array of 3D shapes through significant deformation without fabricating new devices or undergoing lengthy processes of material rearrangement and/or resetting. One such example, in particular for 3D deformation, involves combining an elastomeric kirigami sheet with an endoskeleton crafted from a low-melting-point metal and a layer of heating material [3]. Upon inflation, the kirigami elastomer deforms into a pre-programmed 3D shape, while on-demand heating and cooling of the endoskeleton enable the selection of specific deformations. This approach utilizes pre-programming coupled with analogue actuation (inflation) to achieve reconfigurability. However, it is constrained to morph in one direction, making it challenging to achieve complex 3D shapes.

Another strategy utilises 3D-printed soft material embedded with ferromagnetic nanoparticles, which can be deformed with external magnets [5]. In particular, selective polarization of these particles creates a bias for deformation which upon activation of an external magnetic field, causes the kirigami sheets to buckle out of the plane, creating a pre-programmed 3D shape. However, this approach does not achieve full reconfigurability as it only allows the material to morph into one target shape per part. Remarkably, the work from Chen and collaborators [9] came a step closer to develop an example of reconfigurable metamaterial by utilising magnetic actuation to switch the stiffness of the cells of a 2D lattice, marking a significant transition from pre-programmed to re programmable metamaterials.

However, this method faces some limitations as the lattice must be placed in a custom-made instrument to reconfigure the cells. Finally, achieving reconfigurability consists in defining a novel design platform for mechanical metamaterials, which will shift the current paradigm from pre-programming/re-programming to on-the-fly programming.

**Advances in Science and Technology to Meet Challenges**

Being a fairly recent discovery, mechanical metamaterials are difficult to manufacture at a large scale (due to the lack of efficient industrial processes) and are mostly created in research labs via custom fabrication procedures. However, advances in 3D printing (especially via recent techniques that enable printing of a range of soft materials and silicon rubbers) support the development of macroscale metamaterial prototypes at an unprecedented level of complexity, with high potential to achieve miniaturisation. It is now possible to print structures with minimum and/or soluble support, which leaves behind hollow geometries, able to be actuated via pneumatic systems and achieve extreme deformations.

Classic computational modelling approaches, especially Finite Element Analysis and Reduced Order Models, have been extensively employed to unravel the physics of mechanical metamaterials and predict their mechanical behaviour with high accuracy. However, programming devices to obtain target 3D shapes is non-trivial and typically requires the use of optimisation algorithms [2]. Machine Learning methods have recently emerged as tools to solve mechanics problems, including shape-morphing in inflatable systems. However, a machine learning platform to shape morph reconfigurable 3D metamaterials is yet to be developed. For example, to achieve a target shape in a pneumatic cellular metamaterial, one needs to reveal the mapping between the local constraints at each cell and the global deformation of the metamaterial, which is a complex non-linear mapping in an n-dimensional design space. This is the equivalent of solving an inverse non-linear mechanical problem: if I want the material to morph into a target 3D shape, which expansion mode should I constrain in which cell? Recent approaches have used a combination of computational modelling and Machine Learning to reveal such mappings. Specifically, Finite Element Models validated on experiments are then used to calibrate faster Reduced Order Models, which enable efficient scanning of the design space, creating a dataset of solutions to learn from, using for instance, 3D Convolutional Neural Networks.

Lastly, shape morphing and reconfigurability in mechanical metamaterials require control over the single cell deformation, making these devices impossible to operate when the number of cells is large. Various strategies have aimed at simplifying the control of morphing materials through passive elements [10] or embedding intelligence into the system [4]. To achieve input reduction whilst keeping control over the multiple degrees of freedom of the system, new strategies are needed, which require an interdisciplinary effort between engineers, material scientists, and roboticists.

**Concluding Remarks**

In summary, the past few decades have seen a notable increase in research output on shape-morphing metamaterials, highlighting their potential impact across various applications. However, we are still in the early stages of exploration, with much left to uncover.

Mechanical morphing metamaterials are viewed as a representation of embodied intelligence, where design strategies aim to create systems capable of providing computations through their mechanical structure. Additionally, these materials have the potential to offer features like reconfigurability and adaptability, which are crucial for the future development of robots, wearables, and much more.

**Acknowledgements**

Antonio E. Forte would like to acknowledge the financial support of UKRI grant EP/X525571/1 and MR/X035506/1.

# 3 -   Active phononic metamaterials for sensing and signal processing

## G. R. Nash
## Natural Sciences, Harrison Building, North Park Road, The University of Exeter, Exeter, United Kingdom
g.r.nash@exeter.ac.uk

**Status**

The field of phononics has attracted considerable attention in recent years as a method to control acoustic and elastic wave propagation. In this section we focus on one important kind of elastic wave, surface acoustic waves (SAWs), which propagate on the surface of a solid. Although these waves occur over many length scales, here we focus on the waves with micron scale wavelengths that are exploited in a wide range of existing and emerging applications such as filters in telecommunications, microfluidics, and quantum technologies [1]. Since SAW propagation is limited to material surfaces, SAW devices are also sensitive to changes in their environment and can easily be adapted for sensing applications such as mass or gas loading, and chemical or temperature changes.

In analogy to photonics, there has been much recent attention focused on controlling and manipulating SAWs using phononic crystals (PnCs). These can be engineered to exhibit complete bandgaps, prohibiting propagation at particular frequencies (see, for example, Ref. [2]), and can be realized by varying the elastic properties of the media through which they travel. Most SAW phononic crystals have been fabricated by etching holes in the propagation surface, so that the bandgap is due to Bragg scattering caused by the difference in the properties of the air (low acoustic velocity) filling the holes, and the surrounding SAW substrate (high velocity). In this case the bandstructure is principally determined by the geometry of the crystal (period etc), as shown schematically in Figure 1a. More recently, arrays consisting of local resonators (see, for example, Refs. [3] forming a metamaterial have been shown to offer greater design freedom as bandgap frequencies and other characteristics are defined not only by the periodicity of the array, but also by the properties of the resonators themselves.

Both PnCs and local resonator arrays are attractive for realizing new SAW devices with increased performance, or new functionality, compared to existing devices. The ability to dynamically tune the properties of the array, for example the bandgap frequency, using an external stimulus would enable the realization of tunable filters for radar and communications applications. Tuning would also potentially allow SAW beams to be guided and focussed dynamically, controlled, giving the potential for increases chemical or gas sensing using highly resonant cavities that could be switched on and off.

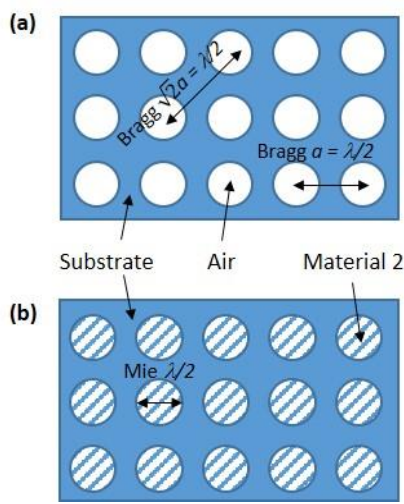

**Figure 1.** Schematic diagram showing plan view of conventional phononic crystal formed from a square lattice, with period $a$, of holes patterned into a substrate filled with (a) air and (b) a second elastic material. In the air filled structure a frequency bandgap will be formed due to Bragg scattering, whereas with the holes filled with a second elastic material Mie scattering will also contribute to the bandstructure.

**Current and Future Challenges**

A simple phononic crystal, formed of air filled holes in a SAW substrate, offers little scope for tunability as the bandstructure is principally determined by the geometry of the patterned array. The most common materials used for SAW devices are lithium niobate and quartz, and the use of these hard materials mean that dynamically changing the phononic crystal geometry is not practicable. It is also hard to dynamically change the elastic properties of either the substrate or the air. One approach for achieving tunability is to consider filling the holes with a second elastic material, as illustrated in Figure 1b, the elastic properties of which can be dynamically varied by some external stimulus. In this case the bandstructure will be determined by a mixture of Mie and Bragg scattering [4], and the characteristics of the phononic crystal are defined by the difference in the properties of the materials used to create it. The mismatch in the acoustic impedances and acoustic velocities govern the frequency, bandwidth and depth (attenuation) of any bandgap formed.

At macroscopic scales many different approaches [5] have been taken to achieve tunabilty, including dynamically changing the geometry, or by the use of photosensitive polymers, but it is not trivial to translate these approaches to the microscopic scale for SAW devices. As well as the challenge in dynamically varying the geometry, the processing technology for typical SAW substrate materials is much less mature that for semiconductor materials, so that just patterning holes, either through reactive ion or focussed ion beam etching, is challenging. Holes can also only be patterned with finite depths, before the sidewalls become sloping, and fabricating large arrays of holes is extremely time consuming. Filling the holes with, for example, a photo- or electro- sensitive polymer adds an extra level of complexity.

Furthermore, it is important to consider the timescales required for tuning. The Rayleigh SAW velocity in lithium niobate is approximately 4000 m/s, so that it typically takes a micro second for the wave to propagate across a device. Switching of the elastic properties of the material filling the holes therefore needs to take place on commensurate time scales. This makes the use of thermal effects difficult, unless local, microscopic heaters could be employed, whereas electrically contacting large number of individual elements would be both practically difficult, as well as potentially limiting the response time due to capacitive effects in large area arrays.

**Advances in Science and Technology to Meet Challenges**

Achieving practicable dynamic tuning in conventional phononic crystals is therefore extremely difficult. Although advances in materials processing techniques will help address some of the technical challenges, such as the time taken to pattern arrays, it is likely that very different approaches will be ultimately required to realise tunability for SAW devices. For example, over the last couple of years topological phononic crystals have been the focus of much research, although there have so far been only a few reports of dynamic tuning (see, for example, Ref. [6]).

Pillared metasurfaces [7] also offer more degrees of freedom than conventional phononic crystals and, for example, Taleb and Darbari [8] studied the use of ZnO pillars for dynamic waveguiding. Their approach made use of the ability to change the elasticity of the ZnO structures by varying their conductivity (by exciting charge carriers using illumination with UV light). Noual *et al* [9] explored the use of plasmonic effects in metal pillars, where light would again act as the external stimulus, although the focus of this work was on the effect of the acoustic waves on the plasmonic resonances, rather than vice-versa.

One way of reducing the complexity of the system is to use approaches where much fewer elements are required. For example, we have previously demonstrated an annular hole array [8] that supports local resonances which create highly attenuating phononic bandgaps. The very large bandgap attenuation achieved means that a smaller numbers of elements is required compared to conventional phononic crystals, making the feasibility of tuning the frequency response of the array easier as any external stimulus would need to affect a small number of elements. Local resonator elements could also be designed to extremely sensitive to an external stimulus, so that can easily be switched from on- to off-resonance, or that different modes are excited within them.

As an example of this we recently showed that using a variant of the annual hole local resonator approach, the square annual hole array shown in Figure 2 [10], we were able to control of SAW phase velocities to values slower and faster than the velocity in an unpatterned substrate; namely, to ~85% and ~130% of the unpatterned SAW velocity, respectively. Rather than using an external stimulus, the reduction or increase in phase velocity originates from the modes to which the SAW can couple at a particular frequency and the displacements induced by local resonators within the array. Such metamaterial approaches open up new avenues for tunability.

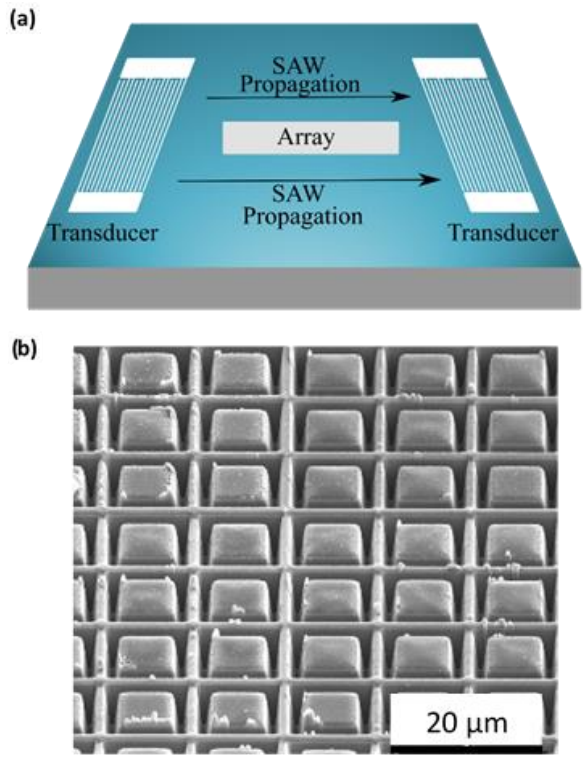

Figure 2. (a) Schematic diagram showing a metamaterial array patterned between the two transducers on a lithium niobate SAW device (b) electron microscope image showing fabricated array. Adapted from [10].

**Concluding Remarks**

Research into SAW devices has been a vibrant and active areas since the invention of the interdigital transducer in the 1960s, which allow SAWs to be directly excited on the surface of a piezoelectric crystal, and there has been considerable recent interest in the use of SAWs for quantum information, microfluidics, and for integration with optomechanical systems. As with photonics the ability to effectively dynamically control the propagation of SAWs would lead to new device concepts and paradigms. Much early research in phononics focused on approaches that were direct analogs of designs used in photonics, and a commonly used approach for SAW phononic crystals is that of etched holes in substrates, where bandgaps are defined by the pitch and filling fraction of periodic arrays through Bragg scattering. However, dynamically tuning the frequency response of such phononic crystals is extremely challenging. Phononic metamaterials consisting of local resonators have been shown to offer greater design freedom, as bandgap frequencies and other characteristics are defined not only by the periodicity of the phononic crystal, but also by the properties of the resonators themselves. These new metamaterial approaches are a promising avenue for the development of ways to dynamically control the propagation of SAWs.

**Acknowledgements**

The author is acknowledges useful discussions with Dr Caroline Pouya, and financial support from the Engineering and Physical Sciences Research Council (EPSRC) of the United Kingdom (A-Meta: A UK-US Collaboration for Active Metamaterials Research, Grant No. EP/W003341/1).

# 4 -   Active acoustic metamaterials

Lawrence Singleton, Felix Langfeldt and Jordan Cheer
Institute of Sound & Vibration Research, University of Southampton, University Road, Southampton, UK
l.r.singleton@soton.ac.uk, f.langfeldt@soton.ac.uk, j.cheer@soton.ac.uk

**Status**

Acoustic metamaterials (AMMs) can achieve very high levels of control at low frequencies using a sub-wavelength distribution of locally resonant unit cells. AMMs can be characterised by a frequency-dependent effective mass density and/or bulk modulus, which can even become negative [1]. However, this behaviour is only exhibited over a small bandwidth. The introduction of active components allows the characteristic response of a metamaterial to be achieved over a broader bandwidth, or to be variable, meaning a system can be switched between different states, or made adaptable to the acoustic environment. It also removes a reliance on resonant structures, as the required behaviour can be driven directly. The possibility for lightweight panels with high sound transmission loss [2]; switchable and steerable acoustic privacy screens; and cloaking [3] make active acoustic metamaterials (AAMMs) an area of significant interest.

The first AMMs with variable response were phononic crystals with variable geometry [4]. Since then, the development of AAMMs with variable low frequency control has been a key objective. In 2009, Baz, proposed the use of a piezoelectric diaphragm to cause a change in the effective stiffness of a constrained fluid, thereby realising a tuneable effective density [5] (see Figure 1(a)). This approach achieved, with a very simple controller, an operational frequency range of 1 kHz [3]. However, negative properties, and therefore spectral bandgaps in transmission, are not achievable with this design. Alternatively, a tuneable, negative effective bulk modulus can be achieved by integrating a piezoelectric diaphragm [6] or traditional loudspeaker [7] into the volume of a Helmholtz resonator (see Figure 1(b)) in order to allow active control of the cavity stiffness. Active control also introduces the possibility for non-reciprocal transmission (see Figure 1(c)) [8], which has advantages in many applications, including privacy screening, and is an area of increasing interest in AAMMs.

Research into AAMMs is still dominated by 1D implementations. The simulation and validation of 2D and 3D setups is not an easy task, but it is key in the development of practically relevant acoustic control solutions. The robustness of AAMMs to factors such as background noise, uncertainty and time-variance should also be thoroughly explored. Manufacturing methods may also require advancement in order to integrate the large numbers of active components required and the associated electrical circuitry involved. In spite of the challenges ahead, the potential advantages of AAMMs means that this area of research is only likely to grow.

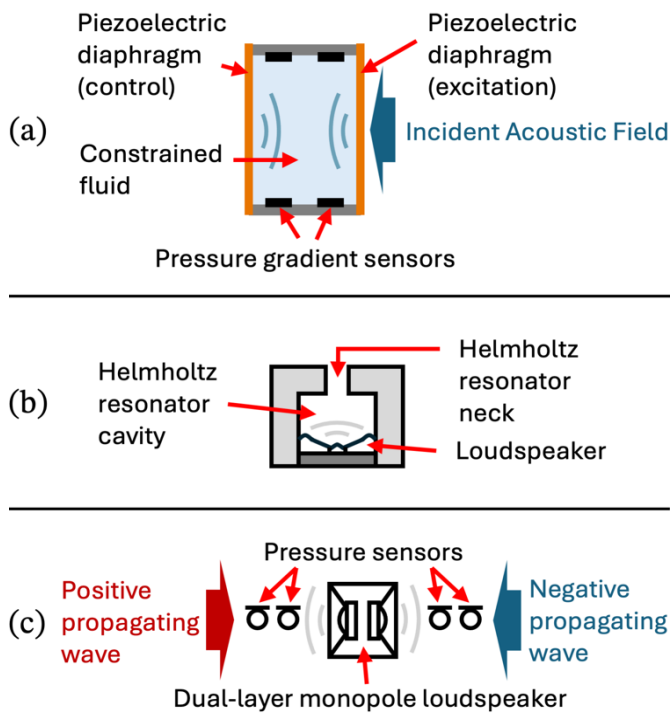

**Figure 1.** Examples of AAMM unit cell designs: a) active effective mass density is achieved in this example by using piezoelectric diaphragms to change the stiffness of a constrained fluid [3]; b) active Helmholtz resonator-based metamaterial for broadband low frequency noise isolation [7]; c) an AAMM capable of non-reciprocal transmission or absorption using wave-based control and a dual-layer monopole loudspeaker [8]

**Current and Future Challenges**

A key challenge for AAMMs is the expansion from a 1D to a 3D environment. In 3D, the interaction with oblique incident waves and nearfield waves must be considered, which is lacking in current research. For example, the distance between two Helmholtz resonators has been demonstrated to dramatically change the transmission loss in a duct [9], so it follows that the performance in a multi-layer Helmholtz resonator-based AAMM will be dependent on incidence angle. AAMMs that actively reinforce the sound field, such as that proposed in [8], will have complex frequency responses due to the interference between acoustic sources. Complex sound fields, impulsive noise, component failure etc. will all have an impact on real-world implementation, and therefore, the robustness in both performance and stability of a system to a wide range of scenarios also needs to be considered. Further, controlling very low frequencies (sub 500 Hz), is likely to require high acoustic output or high levels of deformation depending on the AAMM, therefore, another limitation of real-world sound control using AAMMs is going to be the linear dynamic range of small actuators and loudspeakers. Active control algorithms that deal with output saturation could be used, but these either limit the output, or utilise non-linear controllers with high computational requirements, neither of which are ideal. Where an AAMM drives acoustic pressures directly using loudspeakers or vibroacoustics, the size of the unit cell will also limit low frequency performance due to the volume of air that can be displaced. There is therefore an inherent trade-off between the high frequency limit, dictated by the unit cell size, and the low frequency performance.

For many potential applications (industrial noise, defence, aerospace etc.) the complete system is expected to be large. In commercial terms, perhaps one of the greatest challenges will be to make an AAMM economically viable at those scales. This will require significant investment in manufacturing in order to produce, en masse, elements with integrated sensors, actuators, circuitry, processors etc. How these elements are connected, powered, tested, monitored, repaired, upgraded, even recycled, are all things that need to be considered when scaling up. Additionally, once scaled up, sensors could be shared between unit cells to reduce component requirements or increase performance, or each unit cell could be standalone to simplify the control strategy and manufacturing requirements.

**Advances in Science and Technology to Meet Challenges**

Machine learning-based active noise control methods have been shown to be able to achieve high levels of broadband attenuation of impulsive noise and for non-linear acoustic systems. It is feasible, therefore, that machine learning-based methods may hold the key to stable, robust AAMM realisations. However, the training of a neural network can be time-consuming and complicated, and requires appropriate training data to be available to start with.

As already mentioned, the large-scale manufacturing of metamaterial unit cells with complex electronic componentry and connectivity will require significant investment. Precision processes such as CNC milling and additive manufacturing are time consuming and expensive, and techniques such as injection or blow moulding and hydroforming rely on being able to remove the die(s) from the finished component. In reality, these materials are going to be assembled from component parts for the foreseeable future. However, advances in additive manufacturing have allowed the printing of piezoelectric, magnetic and conductive materials [10]. Although currently in relative infancy, should these processes be perfected, and become economically efficient, they would allow for highly complex networks of sensors and actuators to be laid directly on top of other structures.

Research in AAMMs focusses heavily on novel unit cells and control strategies. Optimisation algorithms have been used as a tool for years to design complex geometries and configurations of resonators to achieve a desired performance, yet the possibility to use optimised design methodologies to maximise linear dynamic range, fatigue resistance, or robustness to uncertainty has so far been somewhat overlooked. For very large arrays of active unit cells, virtual sensing techniques could be instrumental in reducing the number of components required, particularly when multiple sensors are needed for a single unit cell (such as in wave decomposition methods). This would, however, add computational complexity and potentially latency to the system, which would need consideration. With the large number of inputs and outputs to an AAMM, the computational requirements associated with fully centralised control are unlikely to be feasible. Similarly, a fully decentralised system would require a huge number of microcontrollers. Distributed networks have been explored in active control, and as the reliability and speed of wireless networks improves, it is only a matter of time before the technology can support wirelessly, large distributed AAMMs.

**Concluding Remarks**

The development of active acoustic metamaterials promises to completely change what we consider possible in terms of sound control. A variety of approaches have demonstrated tuneable effective material properties and bandgaps, as well as useful non-reciprocal behaviour. This contribution has highlighted significant challenges that lay ahead, primarily in manufacturing and controlling complex multi-sensor, multi-actuator systems on a large scale. With developments in manufacturing and technology poised to overcome these challenges, AAMMs, with infinitely configurable responses, stand to be the future of high performance sound control.

**Acknowledgements**

Lawrence Singleton and Jordan Cheer were partially supported by the Intelligent Structures for Low Noise Environments (ISLNE) EPSRC Prosperity Partnership, UK (EP/S03661X/1). Jordan Cheer was partially supported by the Department of Science, Innovation and Technology (DSIT) Royal Academy of Engineering under the Research Chairs and Senior Research Fellowships programme. Felix Langfeldt was partially supported by the UK's Engineering and Physical Sciences Research Council (EPSRC) UK Acoustics Network Plus EP/V007866/1.

# 5 -    Reconfigurable Metasurfaces and Antennas

Stephen Henthorn
Department of Electronic and Electrical Engineering, University of Sheffield, Sheffield, UK.
s.henthorn@sheffield.ac.uk

**Status**

Microwave metasurfaces are electrically thin structures of conductors and dielectrics which have electromagnetically resonant behaviour in the region of 300MHz to 300GHz. Historically, this research area is closely linked to antennas engineering, as designing materials consisting of passive resonant structures (metamaterials) is not dissimilar from developing resonators to be driven locally (antennas). As such, metasurfaces are currently utilised mostly in antenna enhancement, radar and communications systems. Like other electromagnetic metamaterials they can be used to control the amplitude, polarisation and direction of propagating radiation, both in free space and in other modes such as surface waves. However, due to the use of coherent waves in microwave applications, control of the phase of scattered waves is also very important. By modulating scattered phase across an electrically large area, metasurfaces can redirect microwave energy in predetermined and complex patterns.

Researchers have been investigating how to reconfigure microwave metasurfaces since the 1990s [1]. Three major themes of tuning method have emerged. The most common is device-based tuning, where components are integrated onto the surface which can alter their capacitance, resistance or (more rarely) inductance, changing the resonant behaviour of the structure. Usually this involves soldering on PIN diodes, which act as switches; or varactor diodes, which act as variable capacitors; but alternatives include micro-electromechanical systems (MEMS). These are controlled by an electronic signal, normally a voltage across the device. Second is material-based tuning, where the metasurface structure contains a material which changes its properties – usually permittivity or conductivity – to alter the behaviour of the resonant element. The most common type uses liquid crystals (LCs), such as Kymeta's beamsteering antennas [2], though ferrites, vanadium dioxide and other materials have been used. They are usually tuned using varying strengths of electric fields across the material, though magnetic, acoustic and thermal control is possible. Finally, physical reconfiguration can be used to change the structure of the resonant elements to produce the desired change in behaviour, such as bending, stretching, piezoelectrics and microfluidic control.

Real-time control of an object's behaviour allows several new applications, from simple switches blocking or transmitting propagating waves, to enabling modulation, imaging and even signal processing and non-reciprocal behaviour to be implemented by a surface. If the research challenges can be solved, full and dynamic control of microwave radiation from any object will be possible, allowing metasurfaces to find uses far beyond their current home.

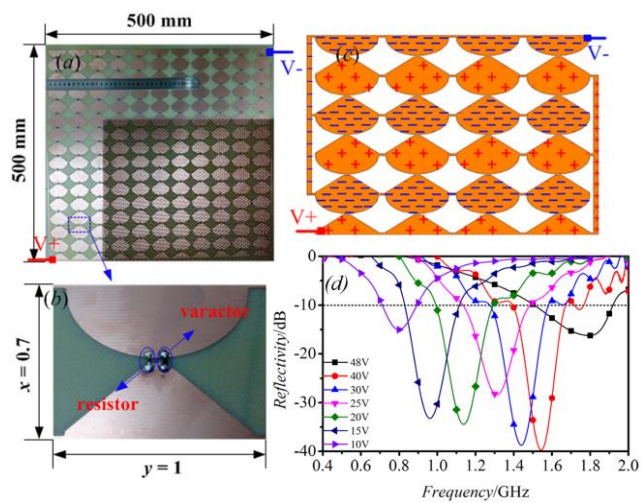

**Figure 1.**  Reconfigurable metasurface absorber using varactor diodes. Modified figure reproduced from IOP (2015) [3]. (a) Photograph, (b) Close-up of unit cell, (c) Diagram showing biasing arrangement to ensure same voltage is applied across all varactor diodes, (d) Measured reflectivity with varying bias voltage.

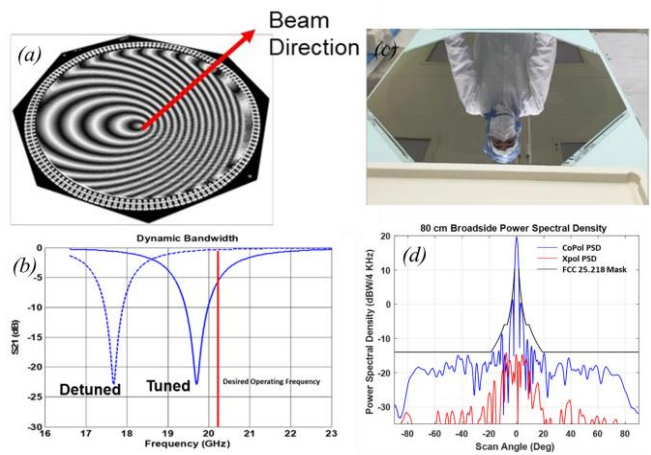

**Figure 2.**  Liquid Crystal tuned reconfigurable metasurface produced by Kymeta [2], (a) Concept of beamsteering using metasurface by switching individual elements on or off, (b) Simulated transmission through the metasurface when either on or off, (c) Photograph of fabricated LC-tuned beamsteering metasurface antenna, (d) Measured copolar and crosspolar radiation pattern from final antenna

## Current and Future Challenges

Almost all microwave metasurfaces rely on resonant behaviour, so the greatest challenge is increasing the bandwidth over which a metasurface can operate. In some cases reconfigurable metasurfaces are a solution to this problem, allowing a single narrowband material to operate over a broader range of frequencies. However, in other applications it introduces issues such as ensuring the phase tuning performance is consistent over a wide range of frequencies.

The rate of switching or tuning is also a consideration. While physical and material tuning are mostly limited by physical properties, it was at first assumed that semiconductor device tuning would enable switching rates in the order of 1GHz. However, to date rates are usually limited to the order of 1MHz [4]. This is because the switching waveforms must be applied on the metasurface to tune the integrated diodes, so when the waveforms have frequency components of a similar order to the resonant frequency of the metasurface structure the material begins to look electrically large to the waveform.

This links closely to the challenge of tuning different meta-atoms of the same material separately, which is desirable in applications such as Reconfigurable Intelligent Surfaces (RIS). Current design techniques model meta-atom behaviour as an infinitely repeating sheet of unit cells, and then assume the meta-atoms maintain this performance even when neighbouring meta-atoms are tuned differently. However, full wave simulation shows this assumption does not always hold due to mutual coupling between meta-atoms. Further, careful design of biasing networks is required to ensure accurate addressing with minimal mutual coupling between control signals to avoid crosstalk.

Similarly, simultaneous yet independent control of different parameters – for example reflected amplitude and phase – is a pressing research challenge. This would allow full control of a material's scattering performance, but decoupling the responses is challenging from both the fundamental physics perspective, and also from the engineering perspective of addressing the properties independently with tuning signals.

Finally, the fabrication and manufacturing challenges must be overcome to enable metasurfaces to be exploited in a wide range of applications. In particular, most reconfigurable metasurfaces either have electronic components soldered onto them, or require complex containing structures to keep in tuneable materials such as LCs. This limits their use in conformal and flexible applications, due to the fragility and lack of integration of these approaches into conventional mass production techniques.

**Advances in Science and Technology to Meet Challenges**

Extending metasurfaces to multiple layers is the usual method of extending their bandwidth, though reconfiguring all layers simultaneously can be challenging. Introducing multiple resonances in a single surface, either near-contiguous to give a broadband effect or deliberately multiband, is also being explored. Whether the reconfigurability of these bands should be completely independent, or completely in sync, depends on the design requirements, and both present their own challenges.

A metasurface element capable of nearly independent amplitude and phase control in a single polarisation has recently been proposed and simulated, though experimental results are not yet published [5]. However, a different approach is to control the aggregate amplitude and phase response of the whole surface rather than individual elements, which has been shown to be possible through control of only the amplitude characteristic of individual elements. Changing the distribution of whether the elements are "on" or "off" allows control of amplitude, phase, polarisation and frequency of the scattered wave [6]. This demonstrates the power of digital coding of reconfigurable metasurfaces, though it requires highly complex control systems to address each element in the surface individually.

Improvement in switching rates of semiconductor tuned metasurfaces is currently a subject of much investigation. Co-design of biasing and control networks and electronics with the structure should become standard practice, along with implementation of good electromagnetic compatibility principles early in the design stage. Optical control by illuminating photodiodes to reconfigure surfaces is another alternative being explored, which minimises the need for electrical biasing networks [7]. Extending this idea further is optically controlling the substrate, such as illuminating a silicon substrate with visible light to change its conductivity [8].

Advanced manufacturing techniques are also being applied to the problem of reconfiguring metasurfaces. This has begun with mechanical reconfiguration, where technologies such as screen printing allows fabrication of metasurfaces on flexible, stretchable or otherwise physically configured structures, which then tune when their shape is changed [9]. Additive manufacture of reconfigurable dielectrics within metasurface structures is another approach, such as using aerosol jet printing to integrate polymer dispersed liquid crystals (PDLCs) within a fully printed structure [10]. This can then be tuned in the same manner as standard LC-based metasurfaces. These methods of integrating reconfigurability into a single fabrication process are important for enabling the uptake of the technology in industrial settings.

**Concluding Remarks**

Reconfigurable microwave metasurfaces are a relatively advanced technology compared with many metamaterials areas, with some early adopters in industry making use of them for tuneable or beamsteering antennas. However, the potential they offer for full control of microwave radiation goes far beyond this. This can be realised by expanding the potential use-cases of metasurfaces by increasing their operating bandwidth and speed of reconfiguration, but should also be accompanied by developing fabrication approaches which can be integrated into existing manufacturing processes.

# 6 -    Reconfigurable photoconductive devices for RF and THz radiation

## Ian R. Hooper and Euan Hendry
## Centre for Metamaterial Research and Innovation, University of Exeter, Exeter, Devon, UK

i.r.hooper@exeter.ac.uk, e.hendry@exeter.ac.uk

**Status**

Metasurface research has turned towards reconfigurable devices in recent years. For frequencies <10s of GHz metasurfaces can incorporate electrical circuitry for tuning [1]. However, this approach becomes increasingly expensive and lossy as the frequency increases, and cannot be implemented for mm-wave and THz frequencies. For these frequencies, utilising optical photoexcitation of charge carriers in semiconductors is one alternative route. Semiconducting regions can be incorporated into complex metasurfaces to gain photocontrol over resistance or capacitance of elements. High mobility semiconductors are necessary to achieve an optimal photoresponse, as one typically wants each absorbed photon to give rise to a large change in conductivity. However, semiconductors which can undergo direct optical transitions such as GaAs or graphene also have short radiative lifetimes, such that conductivity changes are short lived, and modulation requires ultrafast laser excitation. Indirect gap materials such as silicon, on the other hand, can have much longer lifetimes, so that conductivity changes can be "built up" over time using more conventional CW light sources, though at the expense of modulation speed. Here, since they are more suitable for applications, we concentrate on indirect bandgap type modulators.

The first photoconductance modulation of mm-waves was reported in 1980 by Lee et al. [2] where they photoexcited a region of a silicon waveguide through which 94 GHz radiation was propagating. They generated phase shifts of up to 300 degrees/cm, though with very significant absorptive losses, and required an expensive, high-power 50 μJ picosecond pulsed laser system to generate the required carrier densities.

In 1995, a spatially varying photoexcitation was used to tailor the scattering of radiation, with Manasson et al. [3] illuminating a silicon wafer with a periodic pattern using a Xenon flash lamp and a simple binary mask. They achieved approximately 10% diffraction efficiency into the ±1 diffracted orders. It also appears that this paper was the first to identify the need for reducing charge carrier recombination at the surfaces of the wafer (surface passivation) to increase the charge carrier lifetime and maximise the photo-induced carrier density for a given photoexcitation intensity.

In 2012, Gallagher et al. [4] demonstrated a tuneable beam-steerer for 94 GHz radiation by photoexciting a surface passivated Si wafer with a Fresnel zone plate pattern using a commercial projector. They also identified the role of surface passivation in maximising modulation efficiency, and in 2019 Hooper et al. [5] used state-of-the-art surface passivation via atomic layer deposited $Al_2O_3$ to increase the charge carrier lifetime to 10s of ms, enabling modulation depths of >90% of mm-wave radiation with approximately 10 W/m$^2$ - 1/100th of strong daylight - see Fig. 1.

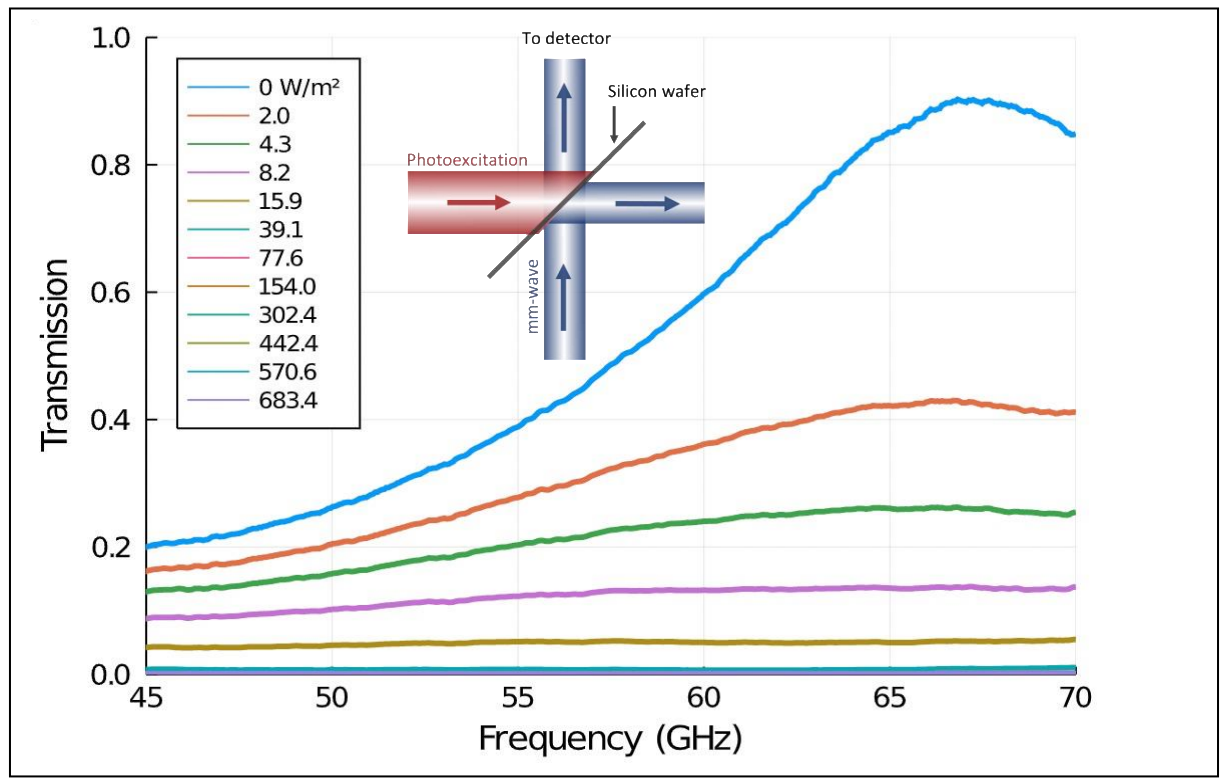

**Figure 1.** The transmission as a function of frequency through a 1 kΩ.cm, 675 μm thick, silicon wafer that has had its surfaces passivated with $Al_2O_3$ to increase the charge carrier lifetime. The transmission through the wafer on Fabry Pérot resonance is reduced by >90% when the Si is photoexcited with approximately 10 W/m² of optical light - just 1/100th of strong daylight [5].

**Current and Future Challenges**

Achieving large modulation depths with low optical powers could open the door to efficient reconfigurable mm-wave and THz spatial modulators using relatively cheap imaging optics. Whilst increasing the charge carrier lifetime increases the modulation efficiency, it comes at a cost - the longer the lifetime, the slower the switching speed and the longer the diffusion length, which can approach several millimetres for charge carrier lifetimes of 10s of ms. The required switching speed will be dependent upon the application, whilst longer diffusion lengths will essentially ``blur out'' any spatial patterning of the photoexcited charge carriers.

To overcome this trade-off, one can use resonant inclusions to increase the light-matter interaction, resulting in increased modulation for a given change in photoconductance, but at the cost of narrow-band performance. The first demonstration of using a photoconductive semiconductor in combination with a resonant metasurface to modulate THz radiation appears to be by Padilla et al. in 2006 [6], where they used an array of split-ring resonators on a GaAs substrate (a direct bandgap semiconductor with a carrier lifetime of ~1 ns) and demonstrated ultra-fast modulation of the transmission of THz radiation using a femtosecond laser system. There have been many similar works since.

However, the requirement to use high-power femtosecond lasers to photoexcite short-lifetime direct semiconductors precludes the possibility of developing cheap *spatial* mm-wave / THz modulators. To do so, longer lifetime indirect semiconductors will have to be used, likely in conjunction with metasurfaces to further enhance the modulation. Since relatively low photoexcitation intensities would be required to essentially "turn off" a resonator, one could use cheap optical projection systems to spatially choose which individual elements of the metasurface interact with radiation, and which do not, resulting in an efficient spatial amplitude modulator. Development of such a device is an open challenge, though a distinctly feasible one.

A further challenge will be to develop spatial *phase* control since this would enable much more efficient control of the reflected / transmitted wavefronts. Due to the mobility of the charge carriers in Si, the change in permittivity upon photoexcitation is predominantly in the imaginary part of the permittivity, and thus works well if one wishes to modify the local amplitude of a field. But how can one utilise a change in the imaginary part of the permittivity to engender a change in the local phase? This is also an open challenge, but will be important for developing efficient spatial modulators for beam-steering etc.

**Advances in Science and Technology to Meet Challenges**

While long charge carrier lifetime semiconductors enable efficient modulation, it would be preferable to have *control* of the lifetime, matching it to required modulation rates. Unpassivated indirect bandgap semiconductors typically have charge carrier lifetimes of the order of a few µs and fully passivated wafers have lifetimes of 10s of ms, but achieving intermediate lifetimes, whereby one can tune it to achieve the switching speed one may desire for a given application whilst maximising modulation efficiency, will be crucial. Recently, Hooper et al. [7] demonstrated a method to lithographically pattern a surface passivation layer to tune the carrier lifetime and control the switching speed of a photomodulator, but simpler methods would be preferable. Going further, it may be possible to control charge recombination at the surfaces by applying a potential - essentially driving the charge carriers away from the surface to prevent recombination. This would allow one to dynamically control the lifetime, and thus the switching speed of a photomodulator. However, this is yet to be demonstrated.

From the metasurface perspective, understanding the interactions between elements will be crucial in developing spatial modulators for beam steering, beam forming, etc. Interactions between resonant elements will create a non-local response - modifying a single element will not only alter the local amplitude/phase, it will also alter the response of its neighbours. One might even be able to turn this non-locality to an advantage, where larger changes to a metasurface's response can be achieved through the excitation of fewer elements, i.e. metasurfaces which are incredibly light sensitive. There are also suggestions that utilising non-locality may give additional degrees of freedom in designing the response of metasurfaces [8]. Understanding how to design metasurfaces with a local response (non-interacting elements), or the way in which any non-locality will impact the performance of a device will be crucial. Understanding the impact of spatially modifying the individual resonators of a non-local metasurface could be a fascinating future area of research, and one where the technologies discussed here could be ideally placed to contribute.

**Concluding Remarks**

The combination of metasurfaces with photoconductive semiconductors and spatial photoexcitation appears to be an attractive option for developing reconfigurable mm-wave and THz devices - a technologically important spectral region where there is a current lack of technologies to do so - for beam-steering, focusing, imaging etc. it will require the ability to control the charge carrier lifetime of the semiconductor, and careful design of the resonant elements that make up the metasurface, to enable accurate *local* spatial control of the amplitude (and preferably phase) of reflected/transmitted wavefronts. However, if this can be achieved such devices could be a transformative step forward in the development of next-generation communications systems and mm-wave/THz technologies.

**Acknowledgements**

I.R.H. acknowledges financial support from the Engineering and Physical Sciences Research Council (EPSRC) via the A-Meta project (Grant No. EP/W003341/1). EH acknowledges support from EP/S036466/1, EP/W003341/1 and EP/V047914/1.

# 7 -   Shape-Morphing Electromagnetic Metamaterials

Alex W. Powell[1] and Anton Souslov[2]

[1] Centre for Metamaterial Research and Innovation, Department of Physics, University of Exeter, Stoker Road, Exeter, EX4 4QL, United Kingdom
a.w.powell@exeter.ac.uk

[2] T.C.M. Group, Cavendish Laboratory, University of Cambridge, Cambridge CB3 0HE, United Kingdom
as3546@cam.ac.uk

**Status**

The combination of shape-morphing (SM) structures with electromagnetic (EM) metamaterials (MM) opens the door to many new functionalities for applications across the electromagnetic spectrum. Shape morphing can be broadly grouped into two classes: 1) Structures which undergo a radical, controlled transformation in shape in response to a simple physical stimulus, (e.g., tension along one axis). 2) Structures that significantly change shape in a controlled manner in response to a change in their environment, (e.g., an increase in relative humidity).

Shape-morphing electromagnetic metamaterials are designed to simultaneously exhibit well-controlled mechanical and electromagnetic properties. The first class of SM-MM includes approaches using origami and kirigami, which have already been used to great effect in electromagnetics to fold deployable antennae for satellites[1] (Fig. 1a). This concept is being commercially applied in the UK by Oxford Space Systems. Beyond packability, the ability to change shape provides a mechanism to switch the effect a material has on reflected or transmitted EM radiation, including its polarisation, frequency, directivity, angular momentum etc., as well as producing electromagnetic behaviours that beyond what is achievable using planar structures [2], [3]. The second class of responsive shape-morphing materials enables the creation of electromagnetic MM that change their behaviour in response to their environment[4], with applications in fields as diverse as medicine, sensing, the internet of things, and architecture. This approach also has the potential to open new routes for the rapid fabrication of metamaterial devices[5].

Another key benefit of shape-morphing metamaterials for electromagnetism is their low power consumption. Typical design approaches for phased array antennas and reflective intelligent surfaces require an electronic switch for every element, usually a PIN diode or varactor. These components can introduce significant losses, and the power consumption for each array grows in proportion to the square of each side length, reducing the gain that can be achieved with any such design[6]. Shape morphing structures, whilst slow to reconfigure, only require a very small number of actuators (potentially only one), and as they are generally chip-free will not suffer losses beyond those induced by the materials they are made up of. Furthermore, many of the base materials underlying these reconfigurable structures are soft, allowing for easy actuation at small force (and energy) scales, and a variety of corresponding actuation mechanisms, from static magnetic and electric fields, to pneumatics and electrical motors.

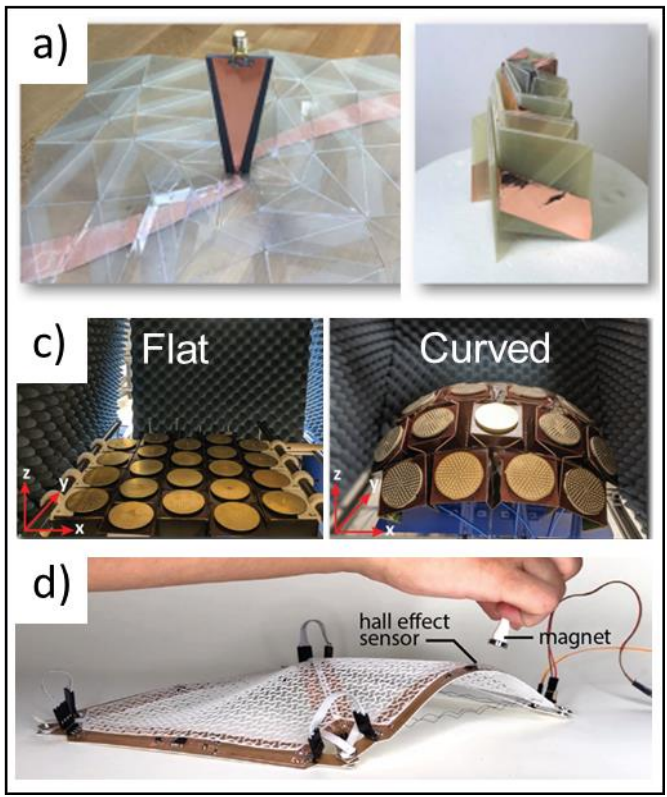

**Figure 1. Examples of some shape morphing structures for electromagnetism: a) An origami-based foldable antenna for satellite applications (An excerpt of the original from [1], licenced under CC BY 4.0). b) An origami microwave imaging array (An excerpt of the original from [2], licenced under CC BY 4.0). c) A shape morphing magnetically responsive structure integrating shape memory alloys, auxetic mechanical metamaterials, magnetic sensors and control systems.**

### Current and Future Challenges

Research into shape-morphing structures for electromagnetism has so far mostly been conducted using quite rudimentary materials and/or methods of actuation. Many designs are still at a 'proof of concept' stage, using readily available (non-optimised) materials, and actuation by hand. Clearly, there is much to be developed before these designs can be taken up in industry. There are three main issues to overcome if SM approaches are to be successful: Materials & fabrication, actuation methods, and tolerances.

**Materials & Fabrication**: A huge range of materials are currently being utilised for SM approaches, from paper to custom designed 3D printable resins[7]. SM-EM structures are often challenging to create, as they typically involve metals and dielectrics, as well as soft and rigid components. A standard method in the lab is to fabricate the various single-material components separately and then assemble them by hand. However, to be commercially scalable, automated approaches such as multi-material additive manufacturing will need to be developed. Additionally, further work is required in order to demonstrate SM properties with materials that are compatible with real applications - taking into account factors such as environmental degradation and the longevity of materials under regular, repeated mechanical deformation.

**Actuation:** Whilst there is a great range of shape-morphing geometries in the literature, many works *do not* combine these with scalable methods of actuation. In future, the field will need to take more of a holistic approach in matching the right geometry, the right material choices, and an appropriate method of actuation to achieve the desired electromagnetic response.

**Tolerances**: When the response of an electromagnetic structure is defined by its physical state, the state needs to be accurately and reliable selected within stringent limits on tolerances. These limits will depend in a large part on the base materials, actuation method, and method of fabrication, and so focusing on the previous two fields will necessarily improve the tolerances of electromagnetic SM-MM. However, the choice of which kind of EM structure is combined with which kind of shape-morphing structure is also important: to maximise performance, the two structures must be combined in a manner where variance due to small mechanical discrepancies is minimised. As electromagnetic coupling between resonant or conductive elements can be highly sensitive to separation distances, it will likely be necessary to design electromagnetic components or unit cells that minimise or avoid EM coupling between elements that move independently or each other.

**Advances in Science and Technology to Meet Challenges**

The greatest challenge for fabricating electromagnetic shape morphing structures is the number of different material types that must be present to achieve the desired functions. Multi-material 3D printing seems to be one of the only available mechanisms to achieve this in a one-step process. Whilst there is not yet a commercially available machine that can create shape-morphing materials with resonant metal elements in one go, research is continually ongoing in this field[8].

New actuation mechanisms are being developed across a range of mechanical metamaterials. Materials can be programmed using robotic components, where arms move or local torques are actuated to locally reconfigure a unit cell[9]. However, many such cell are expensive to fabricate, and difficult to manufacture at scale. Mechanical instability in bistable metamaterials can provide a well-controlled actuation mechanism where one metastable state is quickly reconfigured into a second stable configuration[10]. There is also significant work into materials and methods for magnetic actuation[7], [11], as well as exploration into 3D printable shape memory polymers[12]. An excellent example of combining shape-morphing materials with appropriate actuation is shown in the work of Hwang et al. who develop a SM material of controllable stiffness, and use a combination of elastic tension, with thermal and pneumatic actuation to create SM structures capable of carrying cargo, swimming, and firing water jets[13].

In terms of tolerances, several groups have started to design EM unit cells that are isolated from nearby elements and therefore minimally effected by the mechanical deformations of their neighbours, leading to excellent mechanical tolerance. An example of this was demonstrated recently by Zheng et al[14], who designed a microwave metamaterial lens with mechanically tunable focal length, where each resonant element was surrounded by a copper cover in order to minimise inter-coupling with its neighbours.

Metamaterials researchers could also benefit from collaboration and discussion with researchers in the fields of soft robotics, and human-computer interaction (HCI). These fields are both strongly interested in morphing materials, but often take a more holistic approach, integrating novel materials with actuation methods and control software. Some recent examples of this from HCI are a modular shape-changing interface for prototyping curved surfaces involving software controlled shape-memory alloys[15] (Fig. 1c), and a toolkit embedding pneumatic powered shape-morphing computation into 3D objects[16].

**Concluding Remarks**

Shape-morphing electromagnetic metamaterials present a flourishing subfield of both photonic and mechanical metamaterials. The field has many potential applications, from automatic environmental sensing and response, to lightweight foldable antennas, to low-power reconfigurable reflectarrays and lenses. These materials present many challenges for large-scale fabrication and manufacturing, because they need to combine a shape-morphing scaffold with an actuation mechanism and a strong electromagnetic response, within very specific tolerances. From the perspective of fundamental science, many challenges remain about the limits of multifunctionality towards which these materials can be pushed, and the principles for their effective design. As well as addressing the individual challenges, a more holistic approach is also required going forward for this field to fulfil its exciting potential.

**Acknowledgements**

AWP acknowledges support from a Royal Academy of Engineering Research Fellowship. AS acknowledges the support of the Engineering and Physical Sciences Research Council (EPSRC) through New Investigator Award no. EP/T000961/1.

# 8 - Nanomechanical photonic metamaterials and metadevices

Eric Plum

Optoelectronics Research Centre and Centre for Photonic Metamaterials, University of Southampton, Highfield, Southampton SO17 1BJ, UK.

erp@orc.soton.ac.uk

**Status**

Just like the same carbon atoms may be arranged to form diamond, graphite, graphene, nanotubes or fullerenes with vastly different properties, the characteristics of metamaterials may be controlled by changing the spatial arrangement of their components. The quest to provide optical properties on demand has led to demonstrations of numerous methods for achieving the required relative displacement of the nanoscale components of photonic metamaterials and metadevices. For example, actuation can be controlled by temperature, electrical current, electric field, magnetic field, and light exploiting thermal expansion, resistive heating, as well as electrical, magnetic and optical forces [1]. Compared to natural materials, the properties of nanomechanical photonic metamaterials can be engineered to be many orders of magnitude more sensitive to such external control signals, as illustrated by demonstrations of giant electrostriction [2], electrogyration [3, 4] (Fig. 1), mechanochromism [5] and optical nonlinearity [6,7]. This enables functionalities such as dynamic focusing [8] and optical bistability [9]. Nanomechanical photonic metamaterials typically consist of optically resonant and moving parts that can be made by focused ion beam milling of membranes of nanoscale thickness [1] or electron beam lithography on silicon-on-insulator wafers followed by etching to free the moving parts [10]. Other approaches include kirigami/origami techniques [11], stretching of elastic metasurfaces [12], e.g. to control the focal length [13] and astigmatism [14] of metalenses, and DNA-assembly of resonators actuated by pH changes [15] or DNA-guided structural reconfiguration [16], e.g. to control chirality and its optical manifestations. As mechanical resonance frequencies increase for structures of decreasing size, while a voltage or optical excitation yields a stronger force over a smaller distance, the nanoscale components of nanomechanical photonic metamaterials can deliver devices that are both fast [17] and efficient.

Recent reconfigurable photonic metamaterial demonstrations and observations include sensors, time crystals and noticeable fluctuations of optical properties due to thermal motion.

At non-zero temperatures, all objects are engaged in thermal motion. In typical nanomechanical photonic metamaterials at low pressure, such thermal motion is on the scale of 100 picometers, yields around 0.1% light modulation [18] and can be controlled with light intensity [19] and polarization [20]. This has implications for noise and optimization of nanomechanical photonic metadevices and provides opportunities for the optical characterization of the mechanical properties of the metamaterial's moving parts, including resonance frequencies and damping.

Any quantity – such as light [21] or magnetic field [22] – that influences the actuation or oscillation of a nanomechanical photonic metamaterial may be sensed by observing the metamaterial's optical properties. Dense sensor arrays with sub-diffraction-limited optical readout can be realized by distinguishing the oscillation of different sensing elements based on different mechanical resonance frequencies [23].

While thermal oscillations of nanomechanical photonic metamaterials are random, it was recently observed that optical forces arising from continuous coherent illumination can couple the mechanical oscillators and facilitate a spontaneous phase transition to synchronized motion [24]. This co-called continuous time crystal state opens up opportunities for timing and all-optical modulation.

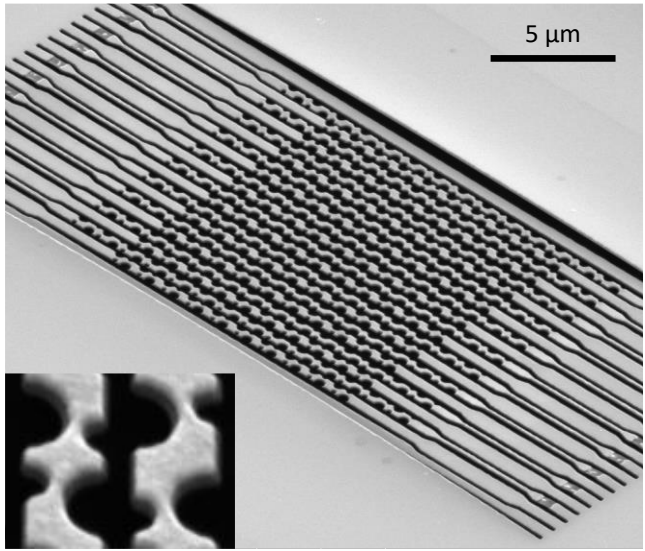

**Figure 1.** Nanomechanical photonic metamaterial exhibiting giant electrogyration. The inset shows the unit cell, which becomes chiral and thus optically active when electric field displaces neighbouring metamaterial nanowires relative to each other [3].

## Current and Future Challenges

While numerous breakthroughs have demonstrated the tremendous potential of nanomechanical photonic metamaterials at the proof-of-concept level, major challenges regarding performance, reliability and integration still need to be addressed.

Indeed, substantial improvements on already impressive results can be expected from optimization of metamolecule designs, material choices and actuation mechanisms. Inverse design using artificial intelligence is already contributing to metamolecule design [25], and can be used to identify designs with higher sensitivity of optical properties to displacement or larger optical forces in response to illumination. High displacement sensitivities can also be expected from perturbation of high-quality-factor optical resonances, for example, Fano resonances in nanomechanical metamaterials made from low loss materials. High displacement sensitivities are particularly important for achieving light modulation with high contrast at high speed, as increasing mechanical resonance frequencies imply stiffer/smaller mechanical resonators, which are more difficult to actuate. Larger displacements of metamaterial components may of course be achieved by larger actuation forces, however, thermal damage limits actuation by heat, electrical current or optical forces, while electrical breakdown and stiction limit actuation by electrostatic forces. For periodic modulation of light, high-quality-factor mechanical resonances can enhance displacement and thus modulation contrast.

The vision of optical properties on demand is yet to be fully realized. To become the ultimate spatial light modulator technology, nanomechanical photonic metamaterials would have to provide complete dynamic control over the wavefront of light with subwavelength resolution. This would require reconfigurable metamolecules of sub-micron size that can control amplitude and phase of light as well as means of addressing them. Significant progress has been made with respect to one-dimensional electrical [26, 27] and optical [23] addressing of nanomechanical metamaterial elements that can control the amplitude and phase of light [28]. Image sensor and spatial light modulator technologies may inspire solutions for two-dimensional electronic addressing, but the much higher pixel density will be a challenge. Another possibility would be diffraction-limited optical actuation of nonlinear metamolecules to control their properties at a longer wavelength with subwavelength resolution in two dimensions.

Regarding mechanically resonant functionalities of nanomechanical photonic metamaterials, recent results have shown that every mechanical resonator tends to have a slightly different resonance frequency, arising from fabrication imperfections and stress. While this may be exploited for optical readout of different sensing elements, it poses a challenge for synchronized and resonantly enhanced driving of nanomechanical metamaterials. Therefore, fabrication of truly identical metamaterial elements – with identical optical properties and identical mechanical resonances – remains a significant challenge.

Arising from the photonic metamaterials community, nanomechanical photonic metamaterials have been developed with a focus on optical properties. Substantial progress can be expected from optimization of designs and material choices with respect to mechanical properties and adoption of best practice from the nano-(opto-)electro-mechanical systems community [29] and related industries. Increasing their involvement may also be expected to accelerate the progression to higher technology readiness levels, which will require high-yield fabrication and long-term reliability.

**Advances in Science and Technology to Meet Challenges**

Fundamentally, progress in nanomechanical photonic metamaterials is a challenge of design, materials and nanofabrication.

Inverse metamaterial design using artificial intelligence can be expected to accelerate progress in nanomechanical photonic metamaterial design by providing means of predicting designs with desirable optical and mechanical properties.

Materials with low optical and/or mechanical losses are needed for high-performance nanomechanical photonic metamaterials. Low mechanical losses enable high quality factor mechanical resonances that can be exploited for resonantly enhanced displacement amplitudes. Plasmonic or high-index dielectric materials with low optical losses are key for achieving high displacement sensitivities, which benefit from high quality factor optical resonances that are limited by absorption losses. Low optical loss materials also contribute to larger optical actuation forces, increased optical damage thresholds and reduced insertion losses. The ongoing search for better plasmonic materials and the growing field of intrinsically low-loss dielectric metamaterials may be expected to drive progress.

Advances in nanofabrication are needed for homogeneous arrays of truly identical mechanical resonators, which currently suffer from inhomogeneous broadening of resonances, and to enable the mass-production of nanomechanical photonic metadevices. While many current reconfigurable photonic metamaterials are made by focused ion beam milling, a high-throughput process will be needed for commercial applications. This could be based on nanoimprint lithography followed by an etching process that frees the moving parts of the metadevice.

**Concluding Remarks**

Nanomechanical photonic metamaterials and metadevices already offer an unprecedented level of dynamic control over light, modulation effects exceeding natural materials by orders of magnitude, a versatile platform for optical sensing and an opportunity to explore fundamental phenomena from thermal motion to time crystals. Substantial improvements on first proof-of-principle devices may be expected from inverse design, advances in nanofabrication and bringing in more expertise from the MEMS/NEMS, opto-mechanics and materials research communities and related industries.

**Acknowledgements**

*This work was supported by the UK's Engineering and Physical Sciences Research Council (grant EP/T02643X/1).*

# 9 -   Active Metamaterials for Thermal Management

Kai Sun[1], C. H. de Groot[2] and Otto L. Muskens[1]

[1] School of Physics and Astronomy, University of Southampton, Southampton UK

k.sun@soton.ac.uk, o.muskens@soton.ac.uk

[2] Electronics and Computer Science, University of Southampton, Southampton UK

chdg@ecs.soton.ac.uk

**Status**

Radiative cooling has a long history for its application in spacecraft thermal management because it is the only means of energy transfer in vacuum.[4] Recently, several groups have proposed significant works on metamaterials-based radiative cooling solutions for terrestrial applications to reduce energy consumption associated with conventional air-conditioning and hence contribute to tackling the net-zero challenge.[5] In case of both space and terrestrial thermal management, the coatings are required to have a low visible and short-wave infrared absorption for rejection of solar radiation and high long-wave infrared (IR) emissivity for radiative cooling through thermal black-body emission, respectively.

Compared with radiative cooling (RC) with a fixed response, active metamaterials (AMMs) offer superior solutions with a tunable IR emissibity, making it desirable in different conditions. Among different materials,[6] vanadium dioxide ($VO_2$) is one of the most popular material for its temperature adapted phase transition around 70ºC, near to room temperature. Recently, the $VO_2/WVO_2$ based radiative cooling solutions (Figure 1) are demonstrated for thermal management in applications of spacecraft,[7] house[2] and windows.[1, 3] These radiative cooling solutions generally fall into two groups, non-transparent and transparent coatings. The former one involves a metallic mirror preferably on flexible substrates, repelling solar radiation and having a tunable IR emissivity. The latter one is like a 'smart' window with a high visible transmittance and also a tunable IR emissivity. Here, the 'smart' RC windows are different from long-term studied 'smart' windows, which focus on visible transmittance tunability through thermochromic $VO_2$ and electrochromic materials. These visible modulated smart windows have been extensively reviewed in recent works[8, 9], and in this work, the focus is given on AMM with RC capability.

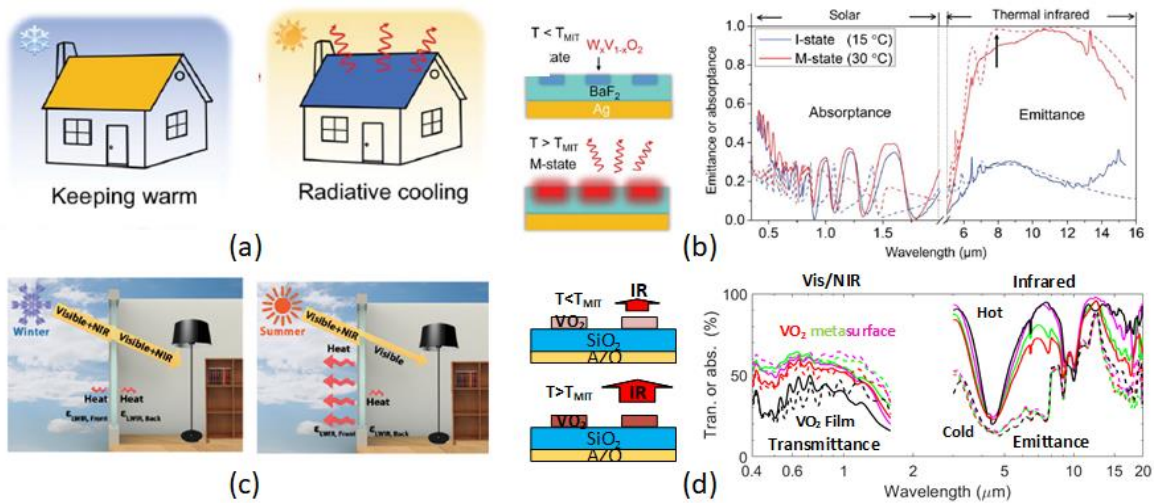

*Figure 1.* (a) smart house roof tunable radiative cooling, (b) W:$VO_2$ based metamaterial radiative cooling solution, (c) smart windows with tunable radiative cooling, and (d) $VO_2$-based metamaterial smart windows. Reproduced from works. [1-3]

**Current and Future Challenges**

For $VO_2$-based thermal management solutions, there are several long-term material challenges. The formation of $VO_2$ involves some high temperature treatment typically above 500ºC, making large scale manufacture over glass or polyimide substrates challenging. In addition, pure $VO_2$ has a transition temperature around 70°C, still above the desirable transition temperature (around 30 °C) in most application cases. The transition temperature has been reported to be tuned down to room temperature through dopant introduction like W.[10] However, the performance is degraded with dopants. Further, the visible absorption of $VO_2$ is quite high and this issue further limits its performance in thermal management. Vanadium oxides (particularly $V_2O_5$) poses serious health concerns for its toxicity and the $VO_2$ nanomaterial's impact on public health is little studied.

_Spacecraft Thermal Management._ In space sector, radiative coating is widely accepted solution. The AMM meets the space desirability for its improvement over Size, Weight, and Power (SWaP). The manufacture of the $VO_2$/W:$VO_2$ metasurface solution involves a costly lithography process and thus posts a scale-up challenge in manufacture cost. Although the space sector has further tolerance in cost, the solution has to be able to sustain strict qualification tests against the harsh space environment, and fully meet technical specification. Further, the space sector is conservative on new solutions with a strong preference over mature solution.

_Architecture Thermal Management._ Radiative cooling coatings and smart windows with radiative cooling are welcomed for energy efficiency. However, its wide implementation is very challenging. Firstly, the construction sector is extremely cost-sensitive for its large volume, leaving Active metamaterial solutions less competitive in value proposition. Secondly, current architecture energy efficiency concepts focus on maximizing thermal insulation of all materials, leaving the cooling and heating to heaters or air conditioners, and there is little thermal conduction through building structures. Thus, radiative cooling on surfaces such as roofs or windows is incompatible with the current system design. Moreover, the $VO_2$ has a visible absorption and demonstrates a brownish-yellow color and the color preference related to cultural background in different regions cannot be ignored. Particularly, the cosmetic appearance matching original design is the top priority for landmark buildings, leaving energy efficiency significantly less leverage. Last, the construction sector adopts whatever needed to meet energy efficiency regulation set by the authority and thus there is no financial incentive for the adaptation of active metamaterials by designers and constructers.

Smart Glazings for electrical vehicles. Compared with the conservative construction sector, electrical car sector welcomes new energy efficient solutions as a practical approach to overcome battery capacity and increase the milage by reducing the energy consumption associated with climate control. Unlike fuel cars, electric cars have adopted a special glazing named low-emissivity window, minimising heat transfer through IR emission. This low-emissivity glazing involves thin metal film and this significantly shield wireless communication, such as 4G, 5G. Recently, car industries are desperate for new solutions as this issue will only worsen in the next auto-driving cars requiring more communications in-cars and between cars. This challenge offers a big opportunity for Active Metamaterial solutions. However, this application space poses further challenges in designing ultra-broadband response control from visible to radio frequency and cost-effective lithography over large-area glazing manufacture.

**Advances in Science and Technology to Meet Challenges**

On $VO_2$ material challenges, the incompatibility of high temperature process can be solved through formation and transfer[2] or low-temperature anneal optimisation works. The $VO_2$ transition temperature can be tuned by W doping, but the optical contrast is affected. Other dopants might be investigated either as W alternative or Co-doping to improve its visible transmittance/absorption and infrared emissivity contrast.[11] The $VO_2$ nanomaterial toxicity is a undeniable concern and needs to be addressed before any wide implementations. Encapsulations of $VO_2$ could be required to protect its overoxidation to $V_2O_5$ in various environments and medical studies should be done on its possible impact to public health.

The cosmetic challenge also needs to be addressed to ensure $VO_2$ radiative cooling coating and windows to meet cultural color preference. One possible solution could be adopt structural color into design.[12] Through a specific stack design, the visible color response can be tailored but this would expectedly further complicate design and compromise its thermal management performance. Additionally, a 'smart' windows with both visible and IR emissivity modulation is also highly desired as a recent work has shown.[13]

Additionally, cost-effective manufacturing is required for bringing $VO_2$-based AMM into various sectors, from architecture to car. It is a grand challenge to form nanostructures over large-area substrates like windows, or flexible substrates, such as polyimide. Here, the authors are optimistic that new technology manufacturing platforms will be integrated when the strong market needs are confirmed. For example, Flat Panel Display (FPD) lithography for LCD production might be modified to do large outputs of nanostructure definitions over large-area windows.[14] For foil-like AMM radiative cooling manufacture, a promising solution is roll-to-roll fabrication techniques, ranging from Nanoimprint lithography, Atomic Layer Deposition, Sputtering and plasma etchings. With marketing needs emerging, these manufacture facilities will be optimized and integrated for scale-up manufacture, leading to a competitive price into market, like photovoltaic panels.

**Concluding Remarks**

Active Metamaterials (AMMs) can provide highly desirable passive thermal management and significantly contribute to net-zero future. We have reviewed the most promising AMM thermal management solution based on temperature adapted vanadium dioxide ($VO_2$) and its challenges over various potential application sectors, space, architecture and electric vehicles. Multisector works from photonic engineering, nanofabrication and medical science will be needed in order to address the challenges and bring its benefits into real life.

**Acknowledgements**

*The author (KS) would like to acknowledge the funding support from Innovate UK ICURe programme on technology commercialization exploration.*

# 10 -  Phase change material based active and reconfigurable optical metasurfaces

Joe Shields[1], Carlota Ruiz De Galarreta[1,2] and C. David Wright[1]

[1] Centre for Metamaterial Research and Innovation, University of Exeter, Devon, UK.

j.shields@exeter.ac.uk, david.wright@exeter.ac.uk

[2] Institute of Materials Science of Barcelona, Universitat Autònoma de Barcelona, 08193 Bellaterra, Barcelona, Spain

cruiz@icmab.es

**Status**

As the state-of-the-art in optical metasurfaces has begun to mature, interest has turned to development of active and reconfigurable optical metasurfaces, the properties of which are not fixed but can be altered and switched post-fabrication. One promising approach in this regard is via the integration of chalcogenide phase-change materials (PCMs) in the metasurface. Chalcogenide PCMs exhibit huge differences in refractive index (real and imaginary parts) between their crystalline and amorphous states, can be switched billions of times between such states (and to fractionally crystallised states) [1] on the (tens of) nanosecond time scales [2], and are stable in such states for years at room temperature. The inclusion of PCMs into metasufaces allows for both a dynamically active optical response, i.e. one that is changed frequently on relatively fast timescales such as in modulation [3, 4], beam steering [5] and display applications [6], and to reconfigurability, where the functionality is switched from time to time for specific needs, e.g. in optical computing [7], optical filtering [8] and reconfigurable lensing [9].

PCMs can be integrated with metasurfaces in a number of ways: by including a continuous PCM layer that acts to modify the resonant behaviour of plasmonic or dielectric meta-atoms [10], by including non-continuous PCM layers into the meta-atoms themselves [1], or by fashioning the meta-atoms entirely out of PCMs [9]. Some PCMs exhibit intrinsic plasmonic properties, so another approach is to laser-write the meta-atoms as crystalline and amorphous regions in such materials [11].

Most reported demonstrations of PCM metasurfaces use the archetypal composition $Ge_2Sb_2Te_5$ (GST), since this material, having been extensively studied for optical and electrical memory applications [12], is well characterised and has a high contrast in refractive index over a wide spectral range. However, the absorption coefficient, k, of GST is non-negligible in the visible and near-infrared regions of the spectrum, and this can introduce unwanted losses in many metasurface configurations. This led to the search for alternative compositions better suited to these spectral regions, with Se-substituted GST, such as $Ge_2Sb_2Se_4Te$ [13], and $Sb_2S_3$ and $Sb_2Se_3$ [14, 15] proving attractive.

Switching PCMs within metasurfaces can be achieved in a number of ways. One approach, demonstrated extensively in the literature [1, 5], is to use external laser pulses to provide the necessary heat to both amorphise the PCM (which requires heating to above its melting temperature) and crystallise it (heating to a temperature below melting optimum for fast crystallisation). Such an ex-situ approach is not, however, likely to be attractive for real-world applications, where in-situ switching is preferable. Such in-situ switching is most practicably achieved using embedded micro-heaters, with approaches using resistive heaters made from metal [16, 17], doped Si [18] and ITO [19] all demonstrated. In some plasmonic PCM metasurface configurations, the metal patterns that provide the plasmonic resonators can double as a resistive heater [10, 17].

In summary, research into PCM metasurfaces is now well advanced: designs incorporating plasmonic, all-dielectric and hybrid plasmonic/dielectric resonators have been demonstrated (see Fig. 1); operation in reflection and transmission modes has been achieved; novel PCM compositions suited to key parts of the spectrum have been developed; a host of interesting metasurface functionalities and potential applications have been showcased; methods for practicable in-situ switching have been reported. However, challenges remain before PCM metasurfaces are suitable for real-world application.

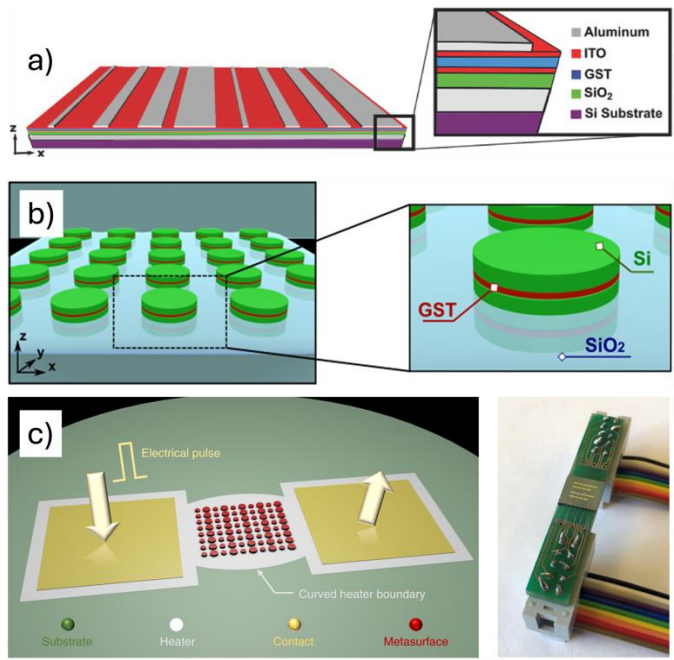

**Figure 1:** a) A typical plasmonic PCM metasurface featuring a continuous layer of PCM. Reproduced with permission from [5]. b) A dielectric PCM metasurface with PCM inclusions within the resonator structures themselves. Reproduced with permission from [1]. c) A PCM metasurface chip with meta-atoms entirely composted of PCM and embedded resistive heaters for in-situ switching. Reproduced with permission from [16].

**Current and Future Challenges**

The structure of PCM-based metasurfaces is more complex than conventional plasmonic (e.g. metal-insulator-metal) or all-dielectric (e.g. Si or Ge nanopillar arrays) metasurfaces, due primarily to a number of constraints arising from the use of PCMs. Firstly, PCMs suffer from oxidation when exposed to air, with effects noticeable after only a few hours of exposure and leading to changes in fundamental PCM properties [20, 21]. The volatile loss of Se and S is also an issue for PCM compositions containing such elements [20]. Thus, practicable designs of PCM metasurfaces should include environmental protection layers, and such layers should not adversely affect the target optical response. Dielectric layers such as $SiO_2$, SiN or $ZnS$-$SiO_2$ borrowed from the realm of optical disk storage [22], $Al_2O_3$ or ITO [5], are commonly used for such protection layers. Secondly, when the PCM metasurface contains plasmonic resonators fabricated from Au or Ag (the usual choices for optimum plasmonic properties), an additional design constraint arises since Au and Ag diffuse readily into chalcogenides, and can alloy with them, degrading the target optical performance. Such diffusion effects are slow at room temperatures, but speed up rapidly at the temperatures required to switch the PCM between its phases. This again means that suitable barrier layers should be included between Au and Ag plasmonic resonators and any PCM layer(s), with $Si_3N_4$ particularly effective in this regard [23, 4]. An alternative approach is to use non-noble metals for the plasmonic resonators, with Al having been shown to provide good optical performance [24]. However, aluminium has a relatively low melting temperature and could suffer from melting and/or structural deformation as a result of the high temperatures required to amorphise most PCM compositions. The amorphisation process brings us to the final, and possibly most testing, design constraint for PCM metasurfaces. Successful amorphisation requires not only heating of the PCM above its melting temperature, but also very rapid cooling to prevent solidification back into the crystal phase. Fast crystallisers such as the archetypal $Ge_2Sb_2Te_5$ composition require cooling rates of around 10 K/ns for successful amorphisation [25], difficult to achieve with very thick PCM layers or meta-atoms (i.e large switching PCM volumes). Fortunately, other PCM compositions such as $Ge_2Sb_2Se_4Te$ and $Sb_2S_3$ (or $Sb_2Se_3$) have crystallisation speeds much lower than that of $Ge_2Sb_2Te_5$ (see e.g. [26, 27, 28]), which relaxes, considerably, thermal design constraints, though at the cost of slower switching capabilities (of the microsecond order).

In addition to the above design challenges, another pressing issue is that of switching the PCM layers within the metasurface. Ex-situ switching using a laser is commonly used in laboratory demonstrations, which is fine for proof-of-concept, but unlikely to be useful for real-world applications, where an in-situ switching approach is much more attractive. Such in-situ switching could be achieved by two main approaches: (i) injecting current into the PCM regions and switching them via Joule heating, or (ii) by embedding resistive microheaters within the metasurface, and heating the PCM regions via thermal conduction. The former approach requires the sandwiching of the PCM regions between electrodes, which again needs to be achieved without significantly degrading the desired optical performance. It is also often difficult to fully switch significant volumes of PCM using this direct electrical switching approach, due to the filamentary nature of the onset of crystallisation [29]. The second approach, using microheaters, is more promising, with several successful demonstrations of reflective and transmissive PCM metasurfaces having been reported using this method [16, 17, 30, 18, 19]. Impressive as such demonstrations are, the metasurfaces used to date have been relatively small (significantly below 1 mm$^2$) and all meta-atoms in the surface were switched in tandem. So, challenges remain in terms of scale-up to larger sizes and/or designs in which not all the meta-atoms should be switched simultaneously. Indeed, the ability to achieve pixel-by-pixel and multilevel (i.e. fractional crystallisation) switching of meta-atoms, or groups of meta-atoms, over such larger areas would pave the way to practicable devices for a wide range of applications including continuous beam steerers, holography, optical angular momentum control, free-space optical computing, metalenses and more.

Another key challenge is that of the endurance, i.e. the number of switching cycles attainable before failure. Electrical PCM memories typically offer between $10^9$ to $10^{12}$ switching cycles [2], while PCM-based re-writable optical disk technologies specified > $10^6$ cycles [25]. Recent studies on blanket $Sb_2Se_3$ films have also shown endurance > $10^6$ cycles [27]. However, the maximum number of cycles demonstrated in PCM metasuface devices to date is typically only a few thousand [31]. This relatively low endurance is most likely due, in large part, to engineering factors, i.e. how well the PCM layers within the metasurface are protected from the oxidation, volatility (for Se and S containing compositions) and diffusion effects previously discussed. Other factors that can reduce endurance include phase segregation (caused by thermo-migration driven by temperature gradients) and cyclic stresses (driven by the significant volume changes of around 4 to 8 percent, depending on composition) that occur upon switching the PCM between phases. The relatively long switching pulses required by slow crystallising materials such as $Ge_2Sb_2Se_4Te_1$ are also likely to be an exacerbating factor, since any thermally degradation effects have longer to take effect (see [20] for a comprehensive review of endurance effects). A key challenge is thus improving the endurance of integrated in-situ switched PCM metasurface devices by suitable engineering optimisation, a non-trivial task but one that has already been addressed in electrical and optical disk PCM memories.

**Advances in Science and Technology to Meet Challenges**

An ideal PCM metasurface would be a device combining high optical efficiency, high endurance and fast in-situ switching, while at the same time being fabricatable over large areas and electrically addressable in a pixel-by-pixel fashion. This, however, represents a formidable engineering task due to multiple, non-trivial and interacting challenges.

If devices with large operating areas and in-situ electrical switching capabilities are to be developed, the use of slow crystallising alloys such as $Ge_2Sb_2Se_4Te_1$, $Sb_2Se_3$ or $Sb_2Se_3$ seems to be, thus far, a viable pathway. This comes with the trade-off of relatively slow switching speeds (typically milliseconds for crystallisation [32]) and lower endurance (currently below 2000 cycles [31]). Improvements might be made, however, by (i) incorporation of suitable low porosity protection layers of sufficient thickness to prevent oxidation and S or Se evaporation [15, 33, 34, 20]), (ii) identification of the optimal pulse switch amplitude and duration for enhanced endurance [20], and (iii) optimisation of the heater design to minimise temperature non-uniformities. Another possible pathway is the exploration of novel alloys having a lower melting temperature and a smaller difference between transition (i.e. crystallisation and melting) temperatures. This should in principle ease the switching process, while reducing the electrical energy required to melt the material. In this context, some novel compositions recently explored for electrical PCM memory applications, such as $NdTe_4$ [35] and $Ge_4Sb_6Te_7$ [36], could be good candidates.

On the other hand, if large endurance and fast switching speeds are required (such as e.g. optical switches for free space interconnects), the use of fast crystallisers, such as GST, will be required. However, special attention should be given to proper thermal design in such cases, in order to achieve the cooling rates required for successful amorphisation. Such thermal constraints can be alleviated via minimisation of the PCM switching volumes, using approaches such as hybrid Silicon/PCM meta-atoms [1, 4].

Finally, most of the available compositions have been specifically synthesised for memory applications, thereby some of their physical properties do not necessarily match what is required by PCM metasurfaces. To this end, the exploration/discovery of novel PCM alloys specifically designed for integration in metasurface architectures is desirable.

## Concluding Remarks

The use of phase-change materials as the active component in tunable optical metasurfaces has shown promise to deliver a wide range of technologically important and novel devices. However, meeting challenges related to increasing the number of switching cycles attainable, to the realisation of large area devices, to pixel-by-pixel addressability and to the desire for fast, multilevel switching - all while retaining the desired optical response - is fundamentally difficult due to the numerous engineering trade-offs at play. Perhaps one approach is to classify, prioritise and address such challenges depending on the application and specifications in mind, much as was done in the past for PCM electrical and optical disk memories. In parallel, scientific efforts should be put on the discovery and synthesis of novel PCM alloys with properties specifically tailored to the needs of PCM metasurface devices, thereby pushing such devices a step closer to practical civil and industrial applications.

## Acknowledgements

J.S and C.D.W acknowledge financial support from the EPSRC via grants EP/W003341/1 and EP/EP/W022931/1. C.R.de G. acknowledges funding from the MSCA Fellowship 101068089.

# 11 - Active graphene-based metamaterials

## Coskun Kocabas and M. Said Ergoktas

Department of Materials, National Graphene Institute, University of Manchester
coskun.kocabas@manchester.ac.uk , muhammedsaid.ergoktas@manchester.ac.uk

**Status**

Active electromagnetic metamaterials require electrically tunable components. Conventional approaches often rely on the tunable capacitance of a reverse-biased diode attached to metallic resonators. However, the frequency response of these diodes is generally limited to microwave frequencies, restricting their use in terahertz (THz) and infrared wavelengths. To extend the operational range into these higher frequencies, alternative tuning mechanism such as electrooptical materials with broader spectral responses, can be utilized. Graphene, the 2-dimensional crystal of carbon, offers unique electro-optical tunability that covers the entire electromagnetic spectrum from the visible to microwave frequencies (Figure 1a) [1]. This multispectral optical activity is derived from controllable interband and intraband electronic transitions within the linear band structure around the Dirac points (see inset in Figure 1b) [2]. This exceptional band structure, which is protected by inversion symmetry of the crystal, results in graphene behaving as a zero-gap semiconductor with atomic thickness and near-zero effective mass for electrons and holes [3]. These properties make graphene an excellent candidate for active metamaterials.

In parallel with the rapid development of metamaterials, the research on active metamaterials has mainly focused on integrating graphene with passive metamaterial structures or patterning graphene to exploit its two-dimensional plasmons. The first approach harnesses graphene's tunable optical conductivity $\sigma(\omega) = \sigma_{inter} + \sigma_{intra}$ as a damping mechanism in the resonator. Shifting the Fermi energy of graphene by electrostatic gating yields tunable optical conductivities due to Pauli blocking (dominant in the visible and near infrared wavelength) and Drude response of electrons leading THz and microwave activity [4]. The second approach using graphene's 2D plasmons, provide remarkably strong light–plasmon coupling at far-infrared wavelengths compared with that in conventional 2DEGs in semiconductors [5,6]. These localised plasmons have been utilised for mid-infrared biosensing and active metadevices. Figure 1d-1f shows three examples of graphene based active metamaterials operating at vastly different wavelengths [7–9]. Due to this multispectral optical activity, graphene-based metamaterials remain a highly active research area with potential applications ranging from THz communications to defence applications. The future research should focus on addressing technological and material challenges to enable new technologies where conventional semiconductors fall short.

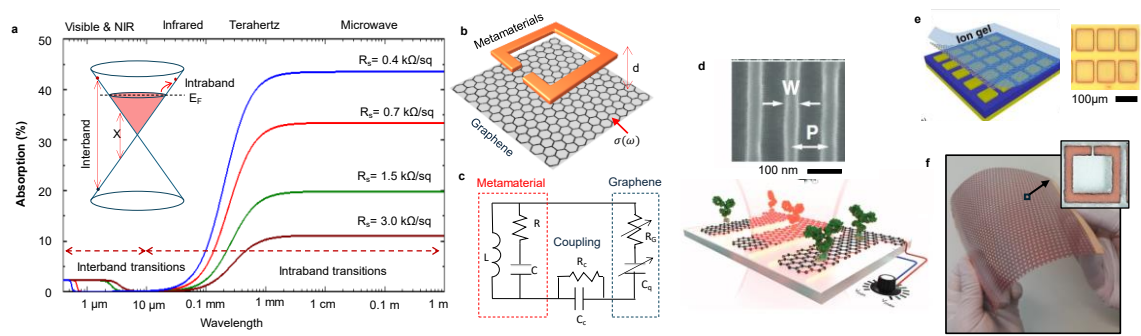

**Figure 1.** (a) Tunable optical properties of single-layer graphene from visible to microwave wavelengths at different charge densities. The inset shows electronic band structure of graphene with the interband and intraband electronic transitions enabling multispectral electro-optical activity of graphene. (b) Schematic representation of a graphene-metamaterial hybrid structure. (c) Equivalent circuit model for the hybrid graphene-metamaterial. Examples of graphene based metadevices illustrating multispectral activity: (d) Patterned single-layer graphene metamaterial used for mid-IR biosensing. (e) THz metasurface consisting of patterned Al mesas covered by single-layer graphene with ion-gel gating. (f) Microwave metamaterial fabricated using large-area graphene capacitors coupled to metallic split-ring resonators.

**Current and Future Challenges**

The pioneering work by Ju et al. demonstrates that plasmon resonances on graphene metamaterials can be tuned over a broad THz frequency range by adjusting the width of the graphene metamaterial and employing *in-situ* electrostatic gating [5]. Unlike plasmons on metals, graphene supports very localized plasmons at the far-IR range. As a rule of thumb, plasmon resonance on graphene is proportional to its Fermi energy. Shifting the plasmon resonance to visible wavelengths requires an extreme Fermi energy shift (>2eV), which is not practical with conventional gating schemes. Biosensing applications of graphene plasmons have been shown to be practical by monitoring infrared reflectivity from the metamaterials. Electrostatic gating provides an additional control parameter to adjust the spectral window of the sensor [7].

The initial demonstrations of tunable intraband transitions in back-gated graphene transistors [10] have been followed by various active hybrid metamaterials, providing new means for controlling the phase and amplitude of THz waves [5,9]. However, the use of transistors has a drawback due to the limited charge density, which constrains the modulation depth. To overcome this limitation, we have introduced an electrolyte gating scheme using graphene capacitors, where two graphene layers gate each other [11]. This geometry eliminates the need for metallic gate electrodes, thereby reducing insertion losses. Additionally, using the concept of coherent perfect absorption, this configuration enhances the tunability of the device, allowing modulation depth >50db.

The operational wavelengths of these devices have been extended to include both infrared and microwave frequencies, each presenting distinct challenges. Achieving high-quality graphene and uniform gating over meter-square areas remains the primary challenge for microwave metamaterials. Meanwhile, infrared wavelengths exhibit lower optical conductivity due to reduced intraband transitions and limited interband transitions. Unintentional doping of graphene also significantly limits its application in infrared technologies. Despite these challenges, there have been successful demonstrations of visible light modulation using ionic liquids as electrolytes [11]. To further enhance the performance of these devices, continued improvements in the gating methods are crucial. All optical switching has been introduced to increase the switching speed to GHz or even THz frequencies [12,13]. Advancing the modulation capabilities into the visible and possibly even into the UV wavelength range is possible, especially where graphene exhibits a Van Hove singularity in its density of states, leading to enhanced absorption of more than 10% for a single layer.

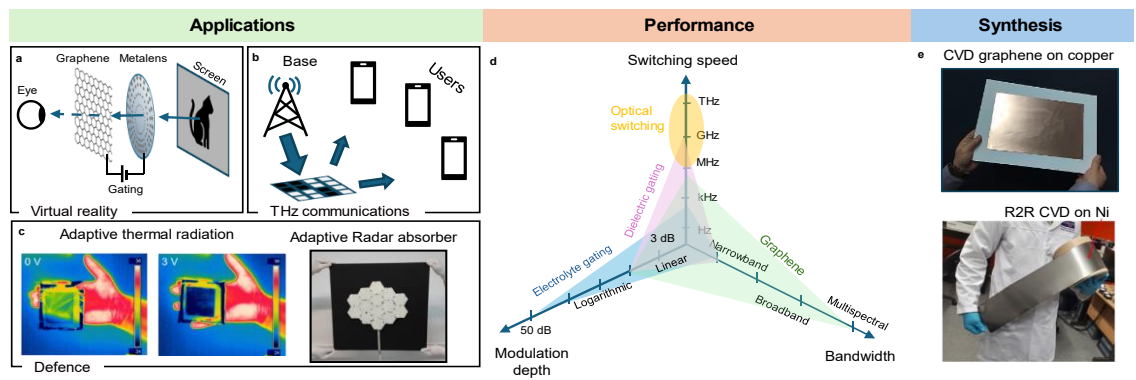

**Figure 2.** Summary of future challenges for active graphene metamaterials: Potential applications include (a) active metalenses for virtual reality devices, (b) intelligent reflective surfaces for THz communications, and (c) smart optical surfaces for adaptive IR camouflage and switchable radar absorbers. (d) Optimization of multiple performance aspects within a single active metadevice. (e) Synthesis and fabrication of high-quality graphene and its integration with conventional microfabrication techniques.

**Advances in Science and Technology to Meet Challenges**

The integration of graphene with gradient metasurfaces could open new possibilities for tunable metalenses, enabling the control of kinetic and geometrical phase gradients at very short propagation distances. Placing a graphene layer in the near field of dielectric or metallic resonators introduces tunable damping, which can be exploited to manipulate light dynamically. This hybrid approach requires large Fermi energy shift to modulate visible light. Demonstration of visible graphene metadevices still remains as a challenge which requires very high-quality graphene and high charge densities >$10^{14}$cm$^{-2}$.

Another promising application is intelligent reconfigurable surfaces, with possible uses for THz communications [14]. Individually addressable pixels containing THz metamaterials and an active graphene layer can control the phase and amplitude of THz reflectivity. These types of intelligent surfaces can be used to steer THz signals to address issues such as low signal strength in THz communications, transmitting THz signals between mowing objects such as cars, satellites. Realization of this technology requires new fabrication schemes to integrate large number of modulators with subwavelength pixels.

Defence applications could benefit from using infrared and microwave graphene metamaterials as adaptive coatings [8,15,16]. For example, thermal radiation from a surface is inherently linked to infrared absorptivity. Controlling the charge density on a graphene surface enables dynamic tunability of infrared absorption and thus thermal emission [17] (Figure 2c). Infrared metamaterials, by pattering graphene or hybrid structures, can enhance the dynamic range of these surfaces. Similarly, the microwave absorption of graphene metamaterials can be adjusted through electrostatic doping. These technologies, however, require a broad spectral response and large dynamic ranges to be effective. Additionally, the integration of such materials into existing defence platform could lead to advancements in stealth capabilities by reducing the detectability of assets via radar and thermal imaging.

These applications have very stringent requirements such as modulation depth, switching speed, and bandwidth. Although there have been many successful demonstrations of these parameters, merging all the requirements into a single device remains a technological challenge. A key limiting factor of graphene metadevices is the gating efficacy, which directly affects the modulation depth. Electrolyte gating offers superior efficiencies; however, its switching speed is limited to sub-kilohertz frequencies. Switching speeds up to GHz frequencies have been demonstrated with dielectric-based devices [18], but their modulation depth and spectral range are inadequate for realistic applications. A hybrid gating method that combines dielectric and ionic technologies could overcome these limitations. This approach could enhance performance across all parameters, enabling the practical implementation of graphene-based devices in a variety of advanced applications, from high-speed telecommunications to dynamic optical systems. Furthermore, new modulation concepts, such as topological switching, could overcome the trade-offs in parameter space. Topological switching leverages the geometrical phases, providing a way to modulate optical reflectivity without the typical compromises [19,20]. Operating a metamaterial near a topological singularity can provide phase modulation up to $4\pi$, depending on the charge of the topological defect, in a short propagation distance.

On the materials side, growing high-quality crystalline graphene over large areas is the key challenge. Over the last decade, there has been significant progress in the growth and commercialization of graphene-based materials. Figure 2e displays commercially available polycrystalline single layer graphene synthesized on copper and multilayer graphene grown on nickel using a roll-to-roll chemical vapor deposition process. Improving quality and reducing costs will be the main focuses for future development of these technologies. Enhancing the scalability of these production methods and achieving higher uniformity and consistency across larger substrates are critical for utilising graphene metamaterials in industries such as defence, telecommunications.

## Concluding Remarks

The exploration and development of graphene-based metamaterials and technologies have shown promising potential across various applications, from advanced defence coatings to intelligent surfaces for telecommunications. The integration of graphene with gradient metasurfaces and other materials enhances the tunability of electromagnetic properties such as THz reflectivity and infrared absorption, which are crucial for applications in stealth technology and dynamic signal management. Despite notable successes in the synthesis and commercialization of graphene, challenges remain in achieving high-quality production on a large scale and merging multiple performance requirements into single devices. The key performance parameter is the maximum achievable charge density which is critically dependent on the gating mechanism. A new gating mechanism together with new modulation concepts would catalyse rapid development for realistic applications.

## Acknowledgements

This research is supported by the European Research Council through ERC-Consolidator grant 682723, Defence Science and Technology Laboratory (DSTLX-1000135951) and EPSRC EP/X027643/1 (ERC PoC grant). Competing interests: C.K. is involved in activities towards the commercialization of graphene-based optical surfaces by SmartIR Ltd.

## 12 - Flexible and conformable photonic metasurfaces

Jianling Xiao, Sebastian A. Schulz and Andrea Di Falco
School of Physics and Astronomy, University of St Andrews, North Haugh, St. Andrews, Fife, KY16 9SS

jx30@st-andrews.ac.uk, sas35@st-andrews.ac.uk, adf10@st-andrews.ac.uk

**Status**

The manipulation of electromagnetic waves is one of – if not the – enabling technologies of the 21st century, laying the foundation for modern communication, imaging and sensing, space observation and fundamental science. In conventional and integrated photonics this manipulation occurs continuously over a finite propagation distance. However, using metasurfaces we can also introduce arbitrary and discontinuous phase changes at an interface, controlled through the arrangement, shape and size of the constituent meta-atoms. Metasurfaces have been explored and demonstrated over a wide frequency range from the visible to the GHz region for imaging, sensing, data storage and encryption, lensing, optical trapping, and cloaking, among others [1].

However, most metasurfaces are fabricated on rigid substrates, which limits their application to a small number of material platforms and form factors. Applying metasurfaces on flexible substrates is an excellent solution to overcome this problem. Due to their flexibility and conformability, they can be transferred to target objects with arbitrary shape and made of arbitrary materials, allowing for easy retrofitting and providing them with additional photonic functionality. Flexible metasurfaces show potential for applications such as patches for new generation telecommunications, wearable photonic skin for high-resolution non-invasive biosensing and detection, soft robotics, and augmented/virtual reality display [1]. Another advantage of flexible metasurfaces is that they can provide more degrees-of-freedom for dynamic manipulation, through e.g. stretching [2].

Over the last decade, the community has made significant progress in the development and use of flexible metasurfaces. Examples include the transfer of a metasurface onto an existing object [3], the introduction of high-efficiency dielectric conformable metasurfaces [2], added functionality such as epsilon-near-zero properties [4] or shape-specific holograms as well as conformable metasurfaces in the mm-wave spectral region [5]. This culminated in the recent demonstrations of tunable lenses [6], shape multiplexed holograms [5] and metasurfaces-based optical traps [7].

However, all the above demonstrations are only scratching the surface of what is possible in flexible metasurfaces, and the benefits coming from the option to couple – or decouple – shape and functionality as one wishes. We will explore this potential and the current challenges in the coming sections.

**Current and Future Challenges**

Until now, the fabrication of flexible metasurfaces has been limited to laboratory settings and sizes in the micrometre or millimetre range. However, mass production and large-scale fabrication with low cost and high efficiency are the key problems that need to be solved, as the current fabrication process is mainly based on electron beam lithography, which requires expensive equipment and is time-consuming. One possible solution enabled through flexible substrates is to change the fabrication process to roll-to-roll UV or nanoimprint lithography. However, this approach still features challenges related to the required metaatom dimensions, compared to typical lithography resolution [8].

The second challenge is conformability. Flexible metasurfaces can be fabricated either directly on a thick film or using a rigid substrate as an intermediate layer for fabrication and transferred to the target object by lift-off [3]. This works well when the object has a simple shape, such as rectangular or cylindrical stacks. However, it would be very challenging to adapt to arbitrary shapes or multi-dimensionally curved shapes in real-world objects. To overcome this issue, one would need stretchable and deformable substrates.

Another challenge is tunability. One of the advantages of flexible metasurfaces is multifunctionality, where different images or information can be encoded into a metasurface and revealed by conforming them to the target's shape [5]. These designs must select the appropriate meta-atoms to provide accurate responses. Controlling the responses of meta-atoms at visible wavelengths is more challenging than at millimeter wavelengths because the meta-atoms size is smaller and because the required accuracy of the mechanical actuation scales with the operating wavelengths.

**Advances in Science and Technology to Meet Challenges**

To achieve flexible metasurfaces in large areas, direct laser writing is a good way to replace UV lithography since it is a maskless method and can be integrated with roll-to-roll fabrication processes. The fabrication resolution is mainly affected by the size of the laser beam, which can be reduced to hundreds of nanometers [9]. It also benefits from low cost, high fabrication speed and can achieve large fabrication areas. Another possible fabrication method is the use of nanoimprinting [10] with high resolution (<100 nm) masks or stamps. For both these methods significant optimization and standardization of the fabrication protocols and potential adaptation of metasurfaces and meta-atom designs is still ongoing, with fast-paced progress.

To meet the challenge of conformability, one possible way is to print the metasurfaces directly on the target object by 3D printing or inkjet printing [1]. A different approach is based on a self-healing strategy which relies on cutting the metasurface into predetermined pieces or simple geometric shapes, such as disks, triangles, or concentric rings, that reduce the local deformation and are optimized to wrap around the target object [1].

Some progress has been made in tackling the tunability problem in the visible range, such as interleaved or segmented holograms which combine different meta-atoms that work with different light properties or revealed conditions [6]. To obtain the real-time response of the meta-atoms, different modulation methods have been explored, such as using electric-induced materials, thermal-varying materials, or using liquid crystals to change the properties of the meta-atoms under different conditions to get different response [6]. The fabrication of such devices on flexible substrates introduces additional complexity that still needs to be resolved.

**Concluding Remarks**

We have summarized the importance and potential applications of flexible and conformable photonic metasurfaces. We have also pointed out the current developments and challenges and discussed the possible solutions to overcome them. Once these are solved, future developments will target materials with desirable properties such as lightweight, low cost, lossless, biocompatible, and degradable, and with a high refractive index, for the next generation of wearable metasurfaces in optical, thermal, mechanical, and acoustic fields, especially for telecommunications, biophotonics, and aerospace applications.

**Acknowledgements**

We acknowledge support from the European Research Council (ERC, Grant Agreement No. 819346) and the EPSRC (EP/X018121/1).

## 13 - Active epsilon-near-zero optical metamaterials

Alexey V. Krasavin and Anatoly V. Zayats
Department of Physics and London Centre for Nanotechnology, King's College London, Strand, WC2R 2LS, United Kingdom
alexey.krasavin@kcl.ac.uk,  a.zayats@kcl.ac.uk

**Status**

The interaction of light with metamaterials is governed by the supported optical modes, particularly their resonant frequencies and dispersion, which are generally determined by the design of the nanostructuring and dielectric permittivity of the constituting media. Among various dispersion regimes of the constituting media or a metamaterial as a whole, the epsilon-near-zero (ENZ) behaviour, which refers to the conditions at which the real part of permittivity approaches zero ($\mathrm{Re}(\varepsilon) \approx 0$), is of particular interest. In this regime, the material exhibits unique optical properties, such as an increased phase velocity and a simultaneously decreased group velocity (slow light effect). The natural materials with the ENZ response are based on the electronic effects in metals and highly-doped semiconductors in the visible and near-infrared, and on phononic resonances in the mid- and far-infrared. The ENZ media have been studied for the enhancement of nonlinear optical effects (both frequency conversion and Kerr-type nonlinear switching), light confinement and guiding in subwavelength waveguides, as well as for metamaterial and metasurface design.

Two approaches can be considered for engineering ENZ metamaterials. In the first one, a thin layer of a natural ENZ material is combined with plasmonic or dielectric nanostructures, resulting in an ENZ metamaterial with properties controlled by resonances and field enhancement provided by the non-ENZ components. In the second one, nanostructures (meta-atoms) made from media with high free-electron concentration are used to realise an effective spectrally-customised ENZ response of an entire metamaterial by shifting its effective plasma frequency through the design of meta-atom geometrical parameters. Using the latter approach, the meta-atom modal structure and the related type of the metamaterial dispersion (elliptical as in usual transparent materials, hyperbolic with different signs of permittivity along the major optical axes, or ENZ) can be engineered in any desired spectral range, providing exquisite control over propagation of electromagnetic waves and their nonlinear interactions, as well as local density of optical states (LDOS) for governing spontaneous emission.

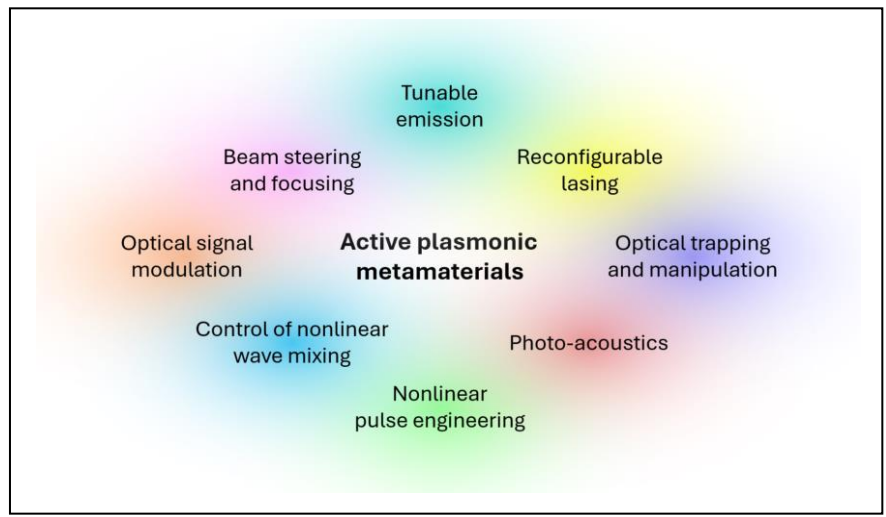

**Figure 1.** Functionalities of active plasmonic metamaterials.

Metallic meta-atoms constituting plasmonic metamaterials support highly-localised plasmonic modes, which additionally provide pronounced field enhancement and have high sensitivity to external stimuli. The latter advantages give plasmonic metamaterials and metasurfaces the ability not only to passively shape spatial, spectral, temporal and polarisation characteristics of reflected, transmitted or emitted light, but also to dynamically control them (Figure 1) [1–3]. Active metasurfaces can be used to steer spontaneous emission of quantum emitters (e.g. quantum dots) [4] and realise reconfigurable lasing [5], as well as photoacoustic modulation and optical trapping/manipulation. These functionalities come at a price of Ohmic losses associated with the plasmonic resonances, which can be partially mitigated by the choice of the plasmonic material in specific spectral ranges. Dynamic metasurfaces can be realised using thermal, opto-mechanical, chemical, magneto-optical, electrical and optical control, utilising various active media, such as molecular, phase-change and 2D materials [6]. In the ENZ regime, electrical or optical stimulation results in unity-scale changes of both real and imaginary parts of the refractive index. Therefore, using ENZ materials as an active medium presents a particular interest in terms of efficiency and/or operation speed, additionally offering easy integration into photonic devices.

**Current and Future Challenges**

Extremely strong nanoscale ENZ-enhanced electro-optic effect in transparent conducting oxides (TCOs) has been used to realise highly-efficient electrically-controlled active metasurfaces for beam steering and switching [1,7] (Fig. 2a). The implementation of the ENZ regime in metamaterials based on traditional (Au, Ag) and new plasmonic materials, such as TCOs and highly doped semiconductors, which spectral position can be customised by the level of doping, leads to the dramatic enhancement of both coherent and incoherent Kerr-type nonlinearities, enabling frequency-mixing and optically-controlled metamaterials and metasurfaces. These nonlinearities can be further boosted by dramatic field enhancement, provided by resonant plasmonic meta-atoms (Fig. 2b) [3,4]. The Kerr-type nonlinear effects, based on excitation of hot electrons upon optical absorption provide optically-induced ultrafast control of permittivity, which can be used to create time-varying optical media, leading to a unique ability to shape the optical field in the frequency space [8]. Alternatively, engineering the metamaterial design and a balance between the metallic and dielectric material components, it is possible to create an artificial nanostructuring-based ENZ response. Using this approach, very strong and fast nonlinear switching has been achieved, while nonlocal (spatial dispersion) effects were used to amplify it further [9] (Fig. 2c). Such nanorod-based anisotropic ENZ metamaterials are also important for spatial and temporal shaping of ultrashort pulses [10] and complex vector beams through the enhanced spin-orbit coupling [11].

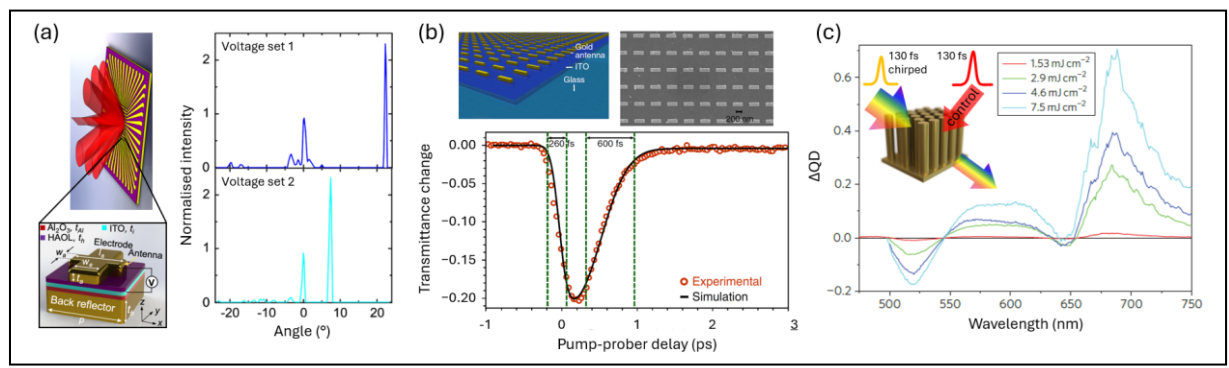

**Figure 2.** (a) Electro-optical steering and (b) all-optical modulation of light signal using ITO-based active metasurfaces. (c) All-optical control of transmission near the ENZ wavelength (around 700 nm) using a nanorod plasmonic metamaterial. (a) Adopted with permission from Ref. 1. Copyright 2020 American Chemical Society. (b) Adopted with permission from Ref. 2. Copyright 2018 Macmillan Publishers Limited. (c) Adopted with permission from Ref. [9]. Copyright 2011 Macmillan Publishers Limited.

The electro-optical effect in electrically-controlled plasmonic metasurfaces, based on charge accumulation in a screening layer at the boundary of TCOs (ITO, AZO, etc.) and dielectrics under the application of voltage is extremely strong, especially after achieving the ENZ condition [1,7]. At the same time, it exists only in a nanoscale region of the accumulation layer. Thus, one of the major challenges here is to achieve the best possible spatial overlap between the accumulation layer and the plasmonic modes of the meta-atoms (also serving as the electrodes), which can be done by optimising the design of the latter. Additionally, fully-customised engineering of spatial light modulation with metasurfaces requires individual control of the optical response of each meta-atom, which in case of the electro-optical approach needs extensive wiring, leading to complexity, heating and increased electronic time delays. Engineering of the component packaging, such as hierarchic vertical integration and uniting the meta-atoms into electrically-controlled assemblies are the possible approaches to mitigate these problems. From the material prospective, one of the major task here is achieving better quality of the nanoscale TCO layers and integrating their deposition into existing industrial fabrication processes. Same applies to alternative electro-optic media, such as graphene and other 2D materials, which have additional challenges of achieving large-scale monolayers and their patterning.

Optically-induced modulation of the optical response of ENZ metamaterials, based on free-electron nonlinearity can be extremely fast, as it is related to optically-induced changes in the energy distribution of electrons in the conduction band, with the corresponding excitation and relaxation times lying in a fs–ps scale. The ENZ regime also allows to reduce the required powers exploiting slow-light effects. Same applies to parametric nonlinear processes, such as harmonic generation and wave mixing.

The non-negligible optical losses in the ENZ media and metamaterials prevent achieving the ideal ENZ response, which otherwise would provide additional advantages in the flexibility of the optical interaction design by decoupling the electric and magnetic fields (since a light wavelength would become infinite).

**Advances in Science and Technology to Meet Challenges**

Recent advances in nanofabrication have led to flexibility of shaping meta-atoms and organizing their arrangement, which allows improved control over the local electromagnetic fields and their coupling, important for designing the optical properties of metamaterials. Particularly, by engineering of the plasmonic meta-atom geometry, it is possible to tune the frequency of the plasmonic resonance and achieve high local field enhancement at the required wavelength as well as shape the field vectorial distribution around the meta-atom, addressing the anisotropic nonlinear response of an ENZ material [12]. Design of the nanostructures also provides a means to affect the excitation and relaxation rates, improving the temporal characteristics of an active metamaterial [13]. Further search for materials with a tailored free-carrier concentration, defining the ENZ wavelength, and increased non-parabolicity, enhancing the nonlinear Kerr-type response, will be highly beneficial from the material side. Alternatively, one can look at another optical modulation mechanisms, for example ultrafast switching between strong and weak coupling regimes of plasmonic meta-atoms and molecular excitons. Finally, for optically-controlled fully-customisable light wavefronts, new designs and approaches for optical addressing of individual meta-atoms need to be developed. Multiple stacked metasurfaces or specially designed 3D metamaterials are needed to achieve phase-matching in order to realise efficient and directional nonlinear wave-mixing at the subwavelength scales.

To reduce losses, high quality monocrystalline plasmonic materials are needed. Meta-atoms formed from single-crystal ultrathin gold films support extremely localised plasmonic modes, offering a greatly improved spatial match to the electro-optical effect based on carrier accumulation in TCOs [14,7]. It can be also expected that resonant transitions between the quantised electron states in such films will also enhance Kerr-driven switching effects.

Technological advances in the realisation of plasmonic and 2D-material nanostructures with non-trivial shapes are accompanied by the development of new theoretical methods for calculation of their band structure, as well as determining electronic nonlinearities and excitation/relaxation dynamics. AI and ML methods can help with the metasurface design and multi-parameter optimisation for the particular functionality or multi-objective performance [15]. From the application point of view, further progress in the research on time-varying media, e.g. utilising strong non-perturbative Kerr nonlinearity in TCOs in the ENZ regime, promises the realisation of non-reciprocal and space-time-coding digital metasurfaces.

**Concluding Remarks**

Active ENZ metamaterials and metasurfaces which allow for real-time tuning of their optical properties are crucial for enabling dynamic control over reflection, transmission, and absorption of light, its phase, polarisation and directionality, as well as spatio-temporal control of ultrashort light pulses and spin-orbit coupling. Compared to their dielectric-based siblings, plasmonic implementations allow straightforward integration with electronics for modulation and detection of optical signals. Similarly, hot-electron excitation provides ample opportunities for achieving strong and ultrafast nonlinear optical modulation and tuneability. With further progress in the search for new natural ENZ materials with improved nonlinear properties which can be used as a basis for metamaterial design, low-loss plasmonic material innovations and development of scalable manufacturing, potentially based on nano-imprint or electrochemical self-organisation processes, active ENZ metamaterials and metasurfaces can play an important role in quantum technologies, optical trapping and manipulation, bio- and gas sensing, applications in reconfigurable nanophotonics and nonlinear optics, including time-varying media. They represent a very attractive platform to create engineered optical materials with designer linear and nonlinear properties, including time-dependent quadratic nonlinearities [3], which can be implemented in free space as well as integrated photonics, exploiting nonlocal and nonlinear effects and strong tuneability, and potentially paving the way to near-zero-index media. Active ENZ metamaterials are also a promising platform for animate materials, reproducing key behaviour of living systems in adaptivity to and self-awareness of the environment.

**Acknowledgements**

This work was supported by UKRI EPSRC projects EP/W017075/1 and EP/Y015673/1, and ERC iCOMM project (789340).

## 14 - Time-varying metamaterials

Emanuele Galiffi

Photonics Initiative, Advanced Science Research Centre, City University of New York (85 St. Nicholas Terrace, 10027, New York, NY, USA)

egaliffi@gc.cuny.edu

**Status**

Whilst metamaterials have opened a new frontier for wave manipulation, numerous challenges stand in the way of widespread technological applications. Some of these are rooted in the passive nature of time-invariant scatterers. These include fundamental bandwidth limits on absorption, reciprocity, and emission-absorption symmetry (Kirchoff's Law or Radiation), and importantly, the significant dissipative and radiative losses undergone by waves propagating through complex media, as well as the intrinsic link between loss and dispersion via the Kramers-Kronig relations. The derivation of these bounds, however, generally assumes the time-invariance of a relevant medium or scatterer.

To overcome these bounds, an unprecedented amount of attention has recently turned to time-varying metamaterials [1]. Time-varying media rely on energy exchanges with an active environment responsible for manipulating their parameters in time. Excitingly, recent works have pointed out the opportunities enabled by time-varying media to overcome absorption bandwidth limits (see Fig. 1a) [2], to overcome conventional absorption-emission relations [3], to enable new paradigms for the control of thermal [4] and stimulated [5] emission, beam steering [6], photonic gauge field engineering [7], and investigations of analogue Hawking radiation [8], among others [1].

Metamaterials that exploit the temporal dimension enable entirely new wave-scattering paradigms. Among these is the concept of time-reflection, dual to spatial reflection, which occurs when the constitutive properties of a material or metamaterial are abruptly modified on a timescale faster than a single oscillation cycle of the waves propagating through it [9]-[10]. This effect, only recently demonstrated for electromagnetic waves [11] (see Fig. 1b), constitutes the broadband counterpart of the steady-state concept of phase-conjugation, whereby a periodic modulation in the wave impedance of a material at twice the frequency of the probe waves leads to its narrow-band time-reversal. Surprisingly, the realisation of time-reflection does not necessarily require a power source, at least at microwaves, but may even be accomplished by means of passive switches, whose net effect is to reduce the energy in the system via mechanisms still under investigation.

Importantly, time-varying metamaterials can enable surprising energy manipulation and pulse-shaping protocols, as recently achieved at microwaves. This is possible because a sufficiently fast temporal modulation breaks the orthogonality between the set of modes in a system (e.g. opposite-traveling waves in a waveguide), enabling the engineering of effective collisions between electromagnetic pulses which can add or subtract power from one another while preserving the net power flow in the system, enabling the mutual "sculpting" of light with light (see Fig. 1c) [12].

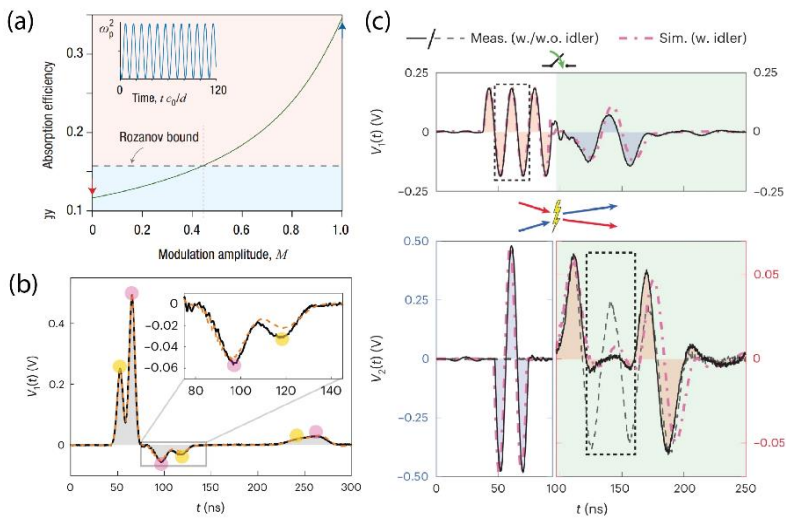

**Figure 1.** (a) Theoretical proposal to use time-modulated structure to enhance the bandwidth limit of a perfect absorber beyond the Rozanov bound [2]. (b) First measurement of photonic time-reflection, in a microwave transmission-line waveguide: the time-reflected pulse is measured at the input voltage probe as a time-reversed signal, with broadened temporal duration due to the change in refractive index [11]. (c) Application of time-reflection to engineer collisions between pulses, enabling distributed instantaneous pulse shaping, whereby a "signal" pulse (top left) and a shorter "idler" pulse (bottom left) are launched from the opposite ends of a waveguide, the latter acting as an "eraser" for the undesired portion of the signal pulse (bottom right, continuous vs dashed black line ) [12].

## Current and Future Challenges

The recent uncovering of these opportunities, combined with the first seminal experiments on time-varying media, have opened several challenges. On the theory front, the insufficiency of the boundary conditions commonly assumed in the literature for temporal inhomogeneities has led to the need for new fundamental electrodynamic considerations to model macroscopic electrodynamics in time-varying media, especially the derivation of generalised boundary conditions which encompass the role of input and output charges and currents, dual to the spatial source terms commonly used for electrodynamic boundary problems [11]. In addition, universal boundary conditions for temporal inhomogeneities in the presence of spatial interfaces, which are expected to encompass additional radiative effects not found in passive media, are still under development.

On the experimental side, efforts are in place to realise these phenomena in higher-frequency regimes, where the new scattering paradigms can play a pivotal role in emerging technologies, in particular wave-based computation, telecommunications and signal processing. For instance, developments in the THz have demonstrated ultrafast, broadband frequency translation based on the pumping of THz waveguides (Fig. 2a) [13]. For instance, on-chip realisation of metamaterials capable of realising time-reflection would project this new concept towards telecommunications applications ranging from time-reversal imaging and intelligent electromagnetic environments to new beamforming strategies (e.g. [12]).

At even higher frequencies, the near-infrared has proved to be a fruitful regime to probe transparent conducting oxides such as Indium Tin Oxide (ITO) under ultrafast pumping conditions. So far, this has led to the observation of large frequency shifts [14], also exemplified by the first realisation of optical time-diffraction (Fig. 2b) [15]. Although other studies have reported promising opportunities for few-femtosecond switching, and even anomalously fast relaxation in ITO, no conclusive theoretical understanding of such ultrafast switching capabilities of ITO has yet been reached, and the observation of optical time-reflection has so far remained elusive, despite the recent use of ultrashort (~5-femtosecond) pump pulses (Fig. 2c-d) [16].

Even more exciting may be the application of these phenomena to quantum light, which may find early implementations with microwave quantum circuits, requiring a synergy between the community working on metamaterials and that working on superconducting quantum interference devices (SQUIDS). Meanwhile, thermal radiation experiments in optically pumped metamaterials may already demonstrate some of the proposals for enhanced manipulation of thermal radiation based on time-modulation, and the breaking of emission-absorption symmetries [3]-[4].

Finally, synergy with key technological sectors in telecommunications and computing will be crucial to establish clear pathways to impact for these emerging concepts towards the development of new practical protocols which leverage time-varying media for signal processing and photonic computation, pulse-shaping, and time-reversal imaging.

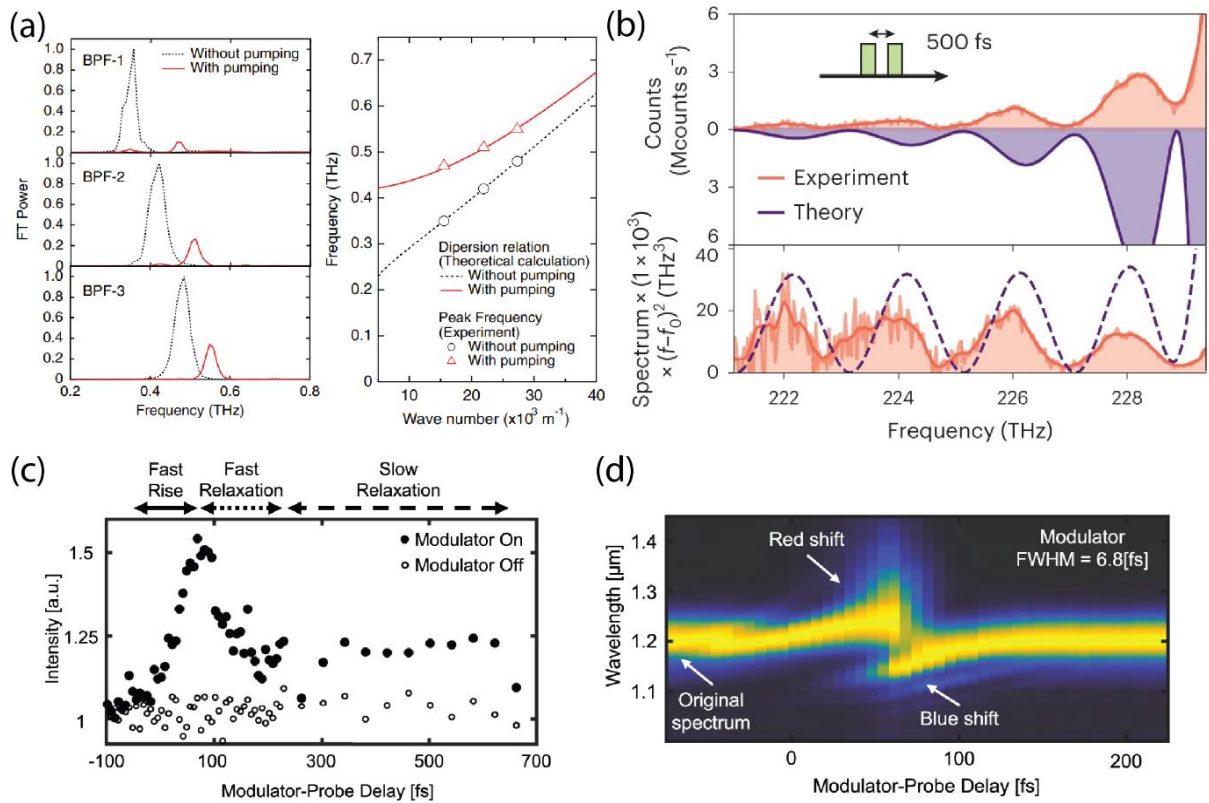

Figure 2. (a) Frequency shifts enabled by optical switching of a terahertz waveguide. The switching occurs as the dispersion relation of the waveguide is abruptly changed by exciting carriers on its top surface, effectively switching it from single to doubly metallised [13]. (b) First demonstration of optical time-diffraction in a thin absorber made of Indium Tin Oxide. The THz-spanning frequency-fringes controlled via the delay between two pump pulses ("slits") witness the opportunity to exploit few-femtosecond switching timescales in transparent conducting oxides [15]. (c) Output intensity and (d) spectrum of a probe pulse transmitted through an ITO film, pumped with ultrashort 6.8-femtosecond pulses [16].

## Advances in Science and Technology to Meet Challenges

At microwaves, the opportunity to exploit new effects such as time-reflection hinges on the possibility to miniaturise and engineer switches that operate faster than the relevant probe timescales. This would be especially exciting if achieved in integrated circuit platforms. However, while this may be achieved through electronics up to a few GHz, hybrid all-optical schemes will need to be devised at higher frequencies, indicating an upcoming crossroad with integrated photonics. An instance of such a multiple-timescale synergy exists in optomechanics, whereby optical pulses are routinely used to manipulate mechanical modes at ~MHz frequencies. Optomechanics may prove a fruitful playground for the realisation of these concepts and to project their applications towards quantum technologies currently under development, as well as a guide towards implementations of multi-physics paradigms for ultrafast time-modulation in other frequency regimes and wave realms.

Moving to the THz, several 2D materials such as graphene exhibit the low carrier densities and high surface-to-volume ratios needed to access giant nonlinearities which could be used to achieve these effects in the mid-infrared via optical pumping. In fact, several groups have recently discovered and measured novel forms of exciton and phonon polaritons in the mid-IR, heralding a rapid expansion of the palette of nanophotonic excitations, whose linear and nonlinear properties offer additional knobs to manipulate nano-optical modes in time [17].

At optical frequencies, although transparent conductive oxides such as ITO exhibit exceptionally large nonlinearities, one primary challenge is the preparation of ultrashort, few-femtosecond (or even attosecond) pulses. Current pumping schemes have largely exploited nonlinearities at frequencies degenerate with the probes, i.e. near the epsilon-near-zero frequency [14]-[16]. However, advantages may be won by engineering and pumping artificial resonances in the visible and ultraviolet to achieve faster modulation for near-IR probes: this is per se a significant challenge, as it hinges on very fine nano-fabrication capabilities. This vision may however prompt further growth in ultraviolet nanophotonics to manipulate optical systems at multiple timescales, combining the delivery of immediate technological advances with a longer-term vision for multi-timescale photonics. An additional fabrication challenge for optical implementations is the non-stoichiometric nature of ITO, which requires individual characterisation of each sample, hindering the reproducibility and standardisation of design, fabrication, and measurement protocols.

Finally, on the theoretical front, the development of efficient computational tools to model the nonequilibrium material dynamics occurring under such ultrashort pumping conditions will constitute a keystone for the design and interpretation of future experiments on time-varying metamaterials. While seminal analytical and computational approaches have recently been put forward [18]-[20], the needed synergy between cutting-edge theory and experiments will require new multi-scale models to bridge the macroscopic electrodynamics and the microscopic nonequilibrium material dynamics spanning multiple timescales in time-varying metamaterials under sub-cycle pumping.

**Concluding Remarks**

To conclude, time-varying metamaterials are unveiling new paradigms for the manipulation of waves and hold the promise to surpass some of the current bounds on passive metamaterials, such as bandwidth limitations and losses. However, despite the unprecedented experimental and theoretical progress of this field in recent years, several challenges lie ahead, which point towards the building of new crossroads with other fields, ranging from ultrafast and attosecond photonics, to out-of-equilibrium light-matter interactions and quantum optics. Hence, addressing these challenges through new cross-disciplinary synergies promises not only to deliver long-term goals specific to the advances of time-varying metamaterials, but also to establish a fertile ground for new cross-disciplinary insights and technological avenues expected to benefit the broader photonics and metamaterials communities in the short term.

**Acknowledgements**

E.G. acknowledges funding by the Simons Foundation through the Simons Society of Fellows programme (855344/EG).